\documentclass[preprint,12pt]{elsarticle}

\usepackage{amssymb}
\usepackage{amsmath}
\usepackage{caption}
\usepackage{subcaption}

\usepackage{xcolor} 

\journal{} %

\begin{document}
\begin{frontmatter}
%
%
%
%

\title{Accelerated multiscale mechanics modeling in a deep learning framework}

%

\author[inst1]{Ashwini Gupta}
\affiliation[inst1]{organization={Department of Civil and Systems Engineering, Johns Hopkins University},
            addressline={\\ 3400 N. Charles Street}, 
            city={Baltimore},
            postcode={21218}, 
            state={MD},
            country={USA}}
            
\author[inst1]{Anindya Bhaduri}

\author[inst1]{Lori Graham-Brady}

\begin{abstract}
Microstructural heterogeneity affects the macro-scale behavior of materials. Conversely, load distribution at the macro-scale changes the microstructural response. These up-scaling and down-scaling relations are often modeled using multiscale finite element (FE) approaches such as FE-squared ($FE^2$). However, $FE^2$ requires numerous calculations at the micro-scale, which often renders this approach intractable. This paper reports an enormously faster machine learning (ML) based approach for multiscale mechanics modeling. The proposed ML-driven multiscale analysis approach uses an ML-model that predicts the local stress tensor fields in a linear elastic fiber-reinforced composite microstructure. This ML-model, specifically a U-Net deep convolutional neural network (CNN), is trained separately to perform the mapping between the spatial arrangement of fibers and the corresponding 2D stress tensor fields. This ML-model provides effective elastic material properties for up-scaling and local stress tensor fields for subsequent down-scaling in a multiscale analysis framework. Several numerical examples demonstrate a substantial reduction in computational cost using the proposed ML-driven approach when compared with the traditional multiscale modeling approaches such as full-scale FE analysis, and homogenization based $FE^2$ analysis. This approach has tremendous potential in efficient multiscale analysis of complex heterogeneous materials, with applications in uncertainty quantification, design, and optimization.
\end{abstract}

\begin{graphicalabstract}
\begin{figure}[h!]
    \centering
    \includegraphics[width=\columnwidth,keepaspectratio]{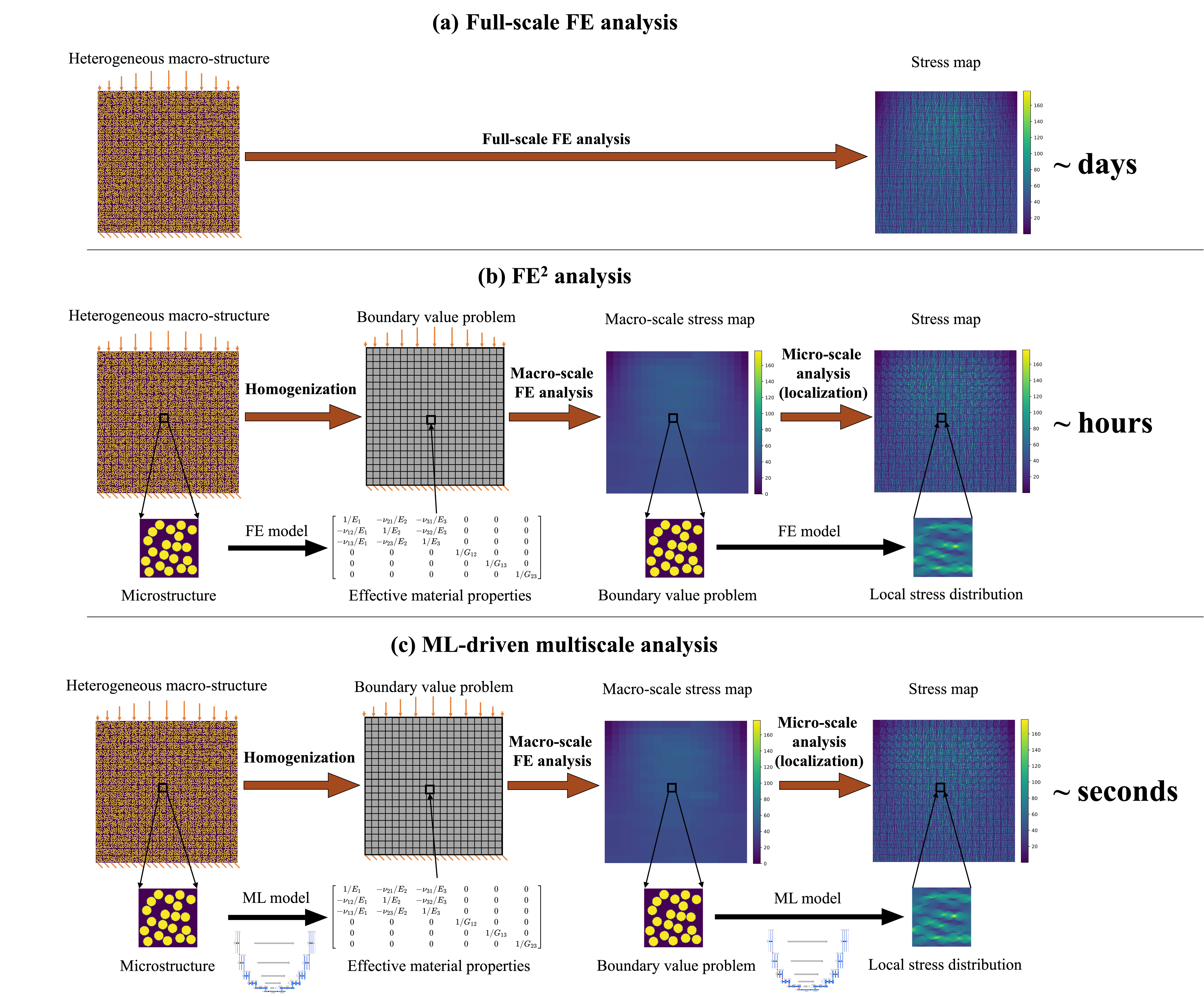}
    \caption{Multiscale mechanics modeling of a heterogeneous macro-structure using three different approaches: (a) full-scale FE analysis, (b) $FE^2$ analysis, and (c) ML-driven multiscale analysis. The full-scale FE analysis is the least efficient, the multiscale FE analysis is parallelizable and, therefore, more efficient, and the ML-driven multiscale analysis is the most efficient as shown by the computation times for each approach to analyze the same number of finite elements.}
    \label{graph_abstract}
\end{figure}
\end{graphicalabstract}



\end{frontmatter}


\section{Introduction}\label{sec:intro}

The micro-scale behavior of heterogeneous materials controls the macroscopic response. Conversely, loading at the macro-scale dictates material behavior at the micro-scale. Therefore, efficient modeling of material behavior and properties across different scales is required for designing materials with targeted properties \cite{olson1997computational, fullwood2010microstructure, llorca2011multiscale}. Conventional multiscale mechanics modeling pertains more to the hierarchical homogenization (upscaling) of materials through effective material properties. However, many critical damage phenomena in composite materials are associated with high micro-scale stresses in the material. It is often computationally infeasible to perform localization (downscaling) to obtain complete micro-scale stresses in a multiscale analysis using physics-based calculations. Recent data-driven machine learning (ML) based approaches have shown tremendous potential in performing efficient micro-scale analysis; however, there is still a need for an ML-based approach that predicts micro-scale stresses in a heterogeneous material under a generalized load state. Here, we develop a novel multiscale analysis approach that performs both upscaling and downscaling, using an ML-model trained on data from a micro-scale analysis. The proposed approach can be applied to a variety of macro-structure shapes and sizes, loadings, and boundary conditions. This ML-driven approach performs multiscale analysis at an unprecedented speed with high accuracy.

\indent Conventional multiscale analysis of heterogeneous structures is based on computational homogenization. An overview of computational homogenization based multiscale methods is provided in \cite{ostoja2006material, geers2010multi, ghosh2011micro, matouvs2017review}. The primary goal of these homogenization based multiscale methods is to relate the effects of material heterogeneity to macroscopic material properties \cite{aboudi1989micromechanical, huet1990application, baxter2000characterization, teferra2018random, wu2018sem}. These works largely focus on the micro to macro transition problems (upscaling), by calculating the average field variables at the macro-scale. The $FE^2$ based multiscale analysis approach presented by Feyel and Chaboche \cite{feyel2000fe2} has applications in both upscaling (homogenization) and downscaling (localization). The two-scale homogenization technique presented by Schr\"{o}der \cite{schroder2014numerical} attaches a representative volume element of the microstructure at each point of the macro-structure. The multiscale strategy proposed by Markovi\^{c} et al.\ \cite{markovic2005multi} associates a portion of the microstructure with each finite element of the macro-structure. In all these high fidelity multiscale approaches, localization using repeated FE analyses requires very large computational effort for practical engineering problems.

\indent Recently, ML-based approaches have shown tremendous potential in design of materials for various applications \cite{butler2018machine, schutt2014represent, mistry2021machine, venturi2020machine, lee2022deep, guo2021artificial, bhaduri2021efficient}. More traditional surrogate models have been applied to homogenization (upscaling) of relevant properties as a replacement to FE analysis \cite{cristianini2000introduction, williams1998prediction, bhaduri2018stochastic, bhaduri2020usefulness, bhaduri2021probabilistic, yang2018deep, pathan2019predictions, bock2019review, mozaffar2019deep, rao2020three, haghighat2020deep, liu2021learning}. There is a growing literature on the use of ML-models to perform efficient upscaling  \cite{liu2016self, wang2018multiscale, liu2019deep, liu2019exploring, liu2021learning, saha2021hierarchical}. These papers discuss upscaling in a multiscale analysis framework, but they do not perform downscaling to calculate high micro-scale stresses developing in the material. Only recently, ML-models have been used to predict local stress fields in heterogeneous material microstructures \cite{yang2019establishing, yang2021end, yang2021deep, rashid2022learning,  sepasdar2022data, bhaduri2022stress, maurizi2022predicting}. However, an ML-model that maps material microstructure to local stress fields under a general load state is still lacking. 

\indent This work proposes a novel multiscale mechanics modeling approach by using an ML-model. The ML-model is trained to predict the stress tensor fields in a random fiber-reinforced composite microstructure resulting from a separate tensile and shear strain loading. This ML-model is capable of predicting micro-scale stresses resulting from a multiaxial strain state by using the superposition principle for a linear elastic material. This pretrained ML-model serves two important purposes in a multiscale finite element modeling context: (i) homogenization (upscaling), to calculate the effective macro-scale properties of each element based on the local element microstructure, and (ii) localization (downscaling), to calculate the sub-element stress distribution corresponding to the element microstructure and the average strains calculated from the macro-scale model. These two phases of the ML-based multiscale model rely on the same training step, mapping microstructure to local stress and/or strain distributions. Therefore, the cost associated with model training is significantly reduced. 

\begin{figure}[h!]
    \centering
    \includegraphics[width=\columnwidth,keepaspectratio]{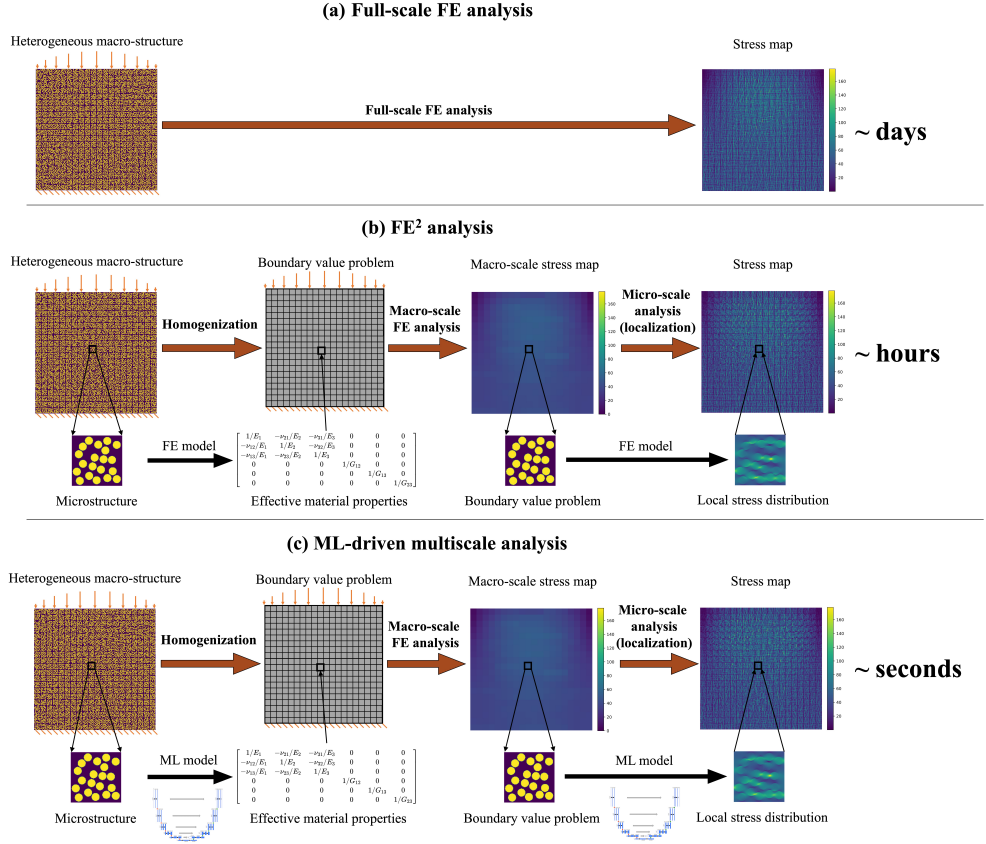}
    \caption{Multiscale mechanics modeling of a heterogeneous macro-structure using three different approaches: (a) full-scale FE analysis, (b) $FE^2$ analysis, and (c) ML-driven multiscale analysis. The full-scale FE analysis is the least efficient, the multiscale FE analysis is parallelizable and, therefore, more efficient, and the ML-driven multiscale analysis is the most efficient as shown by the computation times for each approach to analyze the same number of finite elements.}
    \label{overview}
\end{figure}

Figure~\ref{overview} depicts workflows of the multiscale mechanics modeling of a heterogeneous macro-structure using three different approaches: (a) full-scale FE analysis: the full-scale macro-structure with explicit representation of the microstructure is analyzed using an FE approach, (b) $FE^2$ analysis: a two-scale modeling scheme is implemented by performing micro-scale FE analyses to predict effective macro-scale element properties, macro-scale FE analysis to predict average element stresses/strains, then again implementing micro-scale FE analyses to predict local stresses/strains from the average element strains, and (c) ML-driven multiscale analysis: a two-scale modeling scheme is developed by using ML-model at the micro-scale to predict element properties, followed by a macro-scale FE analysis to predict average stresses/strains, and finally implementing the same ML-model for each element to predict the local stresses/strains from the average element strains. Because of the very large number of degrees of freedom, the full-scale FE analysis is the least efficient. The multiscale FE analysis is parallelizable and, therefore, reduces the computational effort significantly. Because the ML-driven model that predicts the local stress/strain distribution in the microstructure drastically reduces the computational effort relative to FE analysis, the proposed ML-driven multiscale analysis is the most efficient, by orders of magnitude. 

\indent This paper is structured as follows: Section \ref{sec:methodology} describes the proposed ML-driven multiscale mechanics modeling approach. The performance of this approach is evaluated in Section \ref{sec:results}, followed by discussion and conclusions in Section \ref{sec:conclusions}.

\section{Methodology} \label{sec:methodology}
This section describes the ML-driven multiscale mechanics modeling approach. Section \ref{sec:problem_setup} describes the FE model used to generate data for training and evaluating the ML-model. Section \ref{sec:unet} defines the ML-model that maps the microstructure to the local stress fields. Section \ref{sec:effective_material_properties} describes the upscaling (homogenization) approach that uses the ML-model to map a microstructure to its effective properties. Section \ref{sec:multiscale_formulation} presents the ML-driven multiscale analysis approach. This two-scale modeling scheme uses the trained ML-model to perform both upscaling and downscaling, as shown in Figure~\ref{overview}(c).

\subsection{Generation of training data}\label{sec:problem_setup}
\begin{figure}[h!]
    \centering
    \includegraphics[scale=0.4,keepaspectratio]{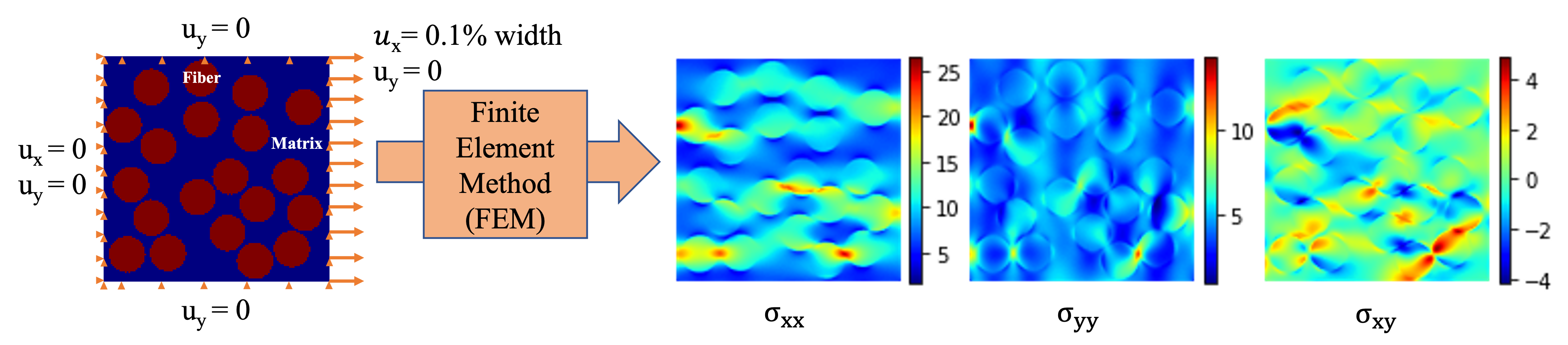}
    (a) Applied tensile strain in $x$-direction
     \includegraphics[scale=0.4,keepaspectratio]{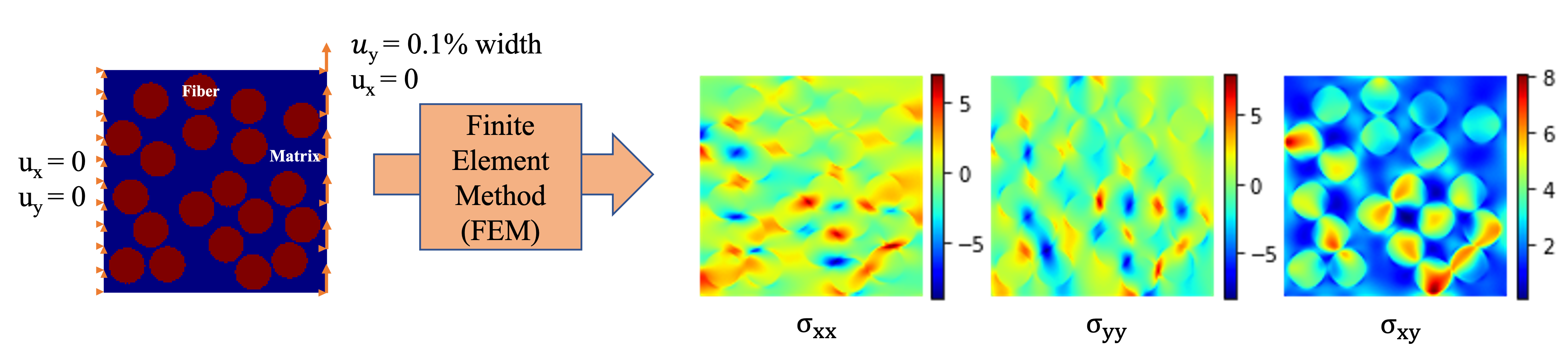}
     (b) Applied shear strain
     \caption{2D composite plate system and the corresponding 2D stress tensor components, as predicted by ABAQUS: (a) under tensile strain in the x-direction, (b) under in-plane shear strain.}
    \label{FEM_model_tensile}
\end{figure}

A fiber-reinforced composite microstructure is analyzed under mechanical tests using FE analysis. The results from this FE analysis are regarded as the ``exact" or true reference solution in this work. A plane strain analysis under an applied displacement boundary condition is performed in ABAQUS \cite{hibbett1998abaqus} to obtain the resulting 2D stress tensor; namely, $\sigma_{xx}$, $\sigma_{yy}$ and $\sigma_{xy}$ distributions in the microstructure. The constituent materials in the composite (fibers and matrix) are assumed linear and elastic with Young's moduli (87 GPa for fiber and 3.2 GPa for matrix) and Poisson's ratios (0.2 for fiber and 0.35 for matrix) taken from earlier work \cite{bhaduri2022stress} for consistency. The fiber/matrix interface is assumed to be perfectly bonded. The microstructure is subjected to two separate displacement boundary conditions corresponding to a tensile and shear strain loading. Two sets of 2D stress tensor fields are obtained from these two simulations in ABAQUS. Figure~\ref{FEM_model_tensile}a indicates the microstructure subjected to a $0.1\%$ tensile strain loading along the x-direction and the resulting 2D stress tensor components. Additionally, Figure~\ref{FEM_model_tensile}b shows the microstructure subjected to a $0.1\%$ in-plane shear strain loading and the resulting 2D stress tensor components. Thus, the FE analyses provide a total of six stress images corresponding to each microstructure image.

\subsubsection{Random microstructure sampling}
\begin{figure}[h!]
    \centering
    \includegraphics[scale=0.35,keepaspectratio]{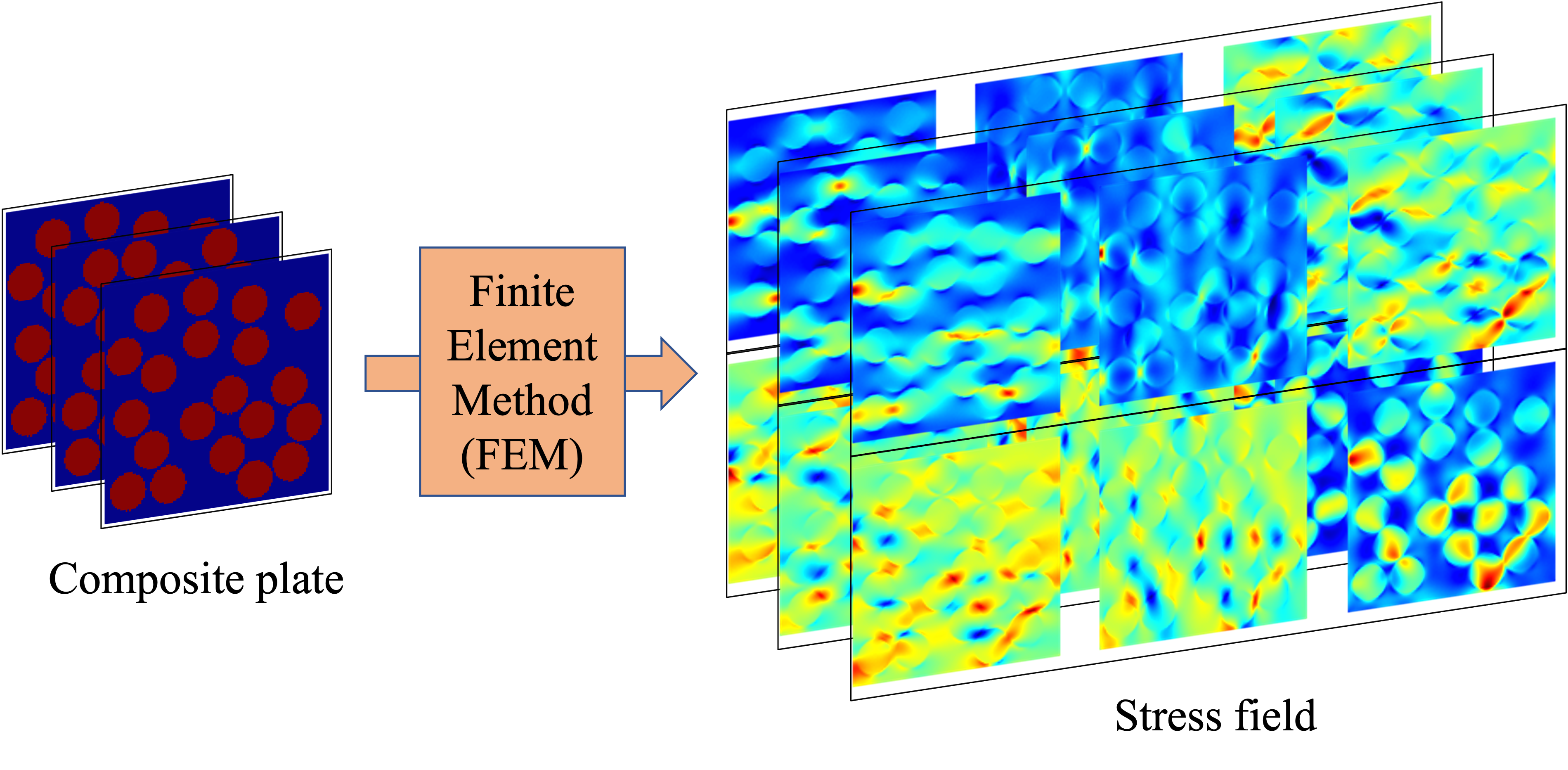}
    \caption{2D composite system composed of circular fibers embedded in a matrix (left) and the corresponding stress tensor field maps resulting from FE analyses under separate tensile and shear strain loading (right).}
    \label{image_stacking}
\end{figure}

A set of 2D images of the fiber-reinforced composite microstructure are generated by randomly varying the spatial location of fibers in the matrix. The fibers are assumed to be circular with a fixed radius, and the fiber volume fraction is kept constant. This results in a collection of microstructures with a random arrangement of $20$ fibers embedded in a matrix. Each microstructure is then subjected to two separate loadings, specifically a $0.1\%$ tensile strain along the x-direction and a $0.1\%$ in-plane shear strain in ABAQUS, to obtain the stress tensor maps corresponding to each loading case as outlined in Section \ref{sec:problem_setup}. The set of random microstructures and the corresponding stress tensor field maps resulting from FE analyses is used as the labeled training data for the ML-model, as shown in Figure~\ref{image_stacking}.

\subsubsection{Data Augmentation} \label{sec:data_augmentation}
\begin{figure}[h!]
    \centering
    \includegraphics[width=0.48\columnwidth,keepaspectratio]{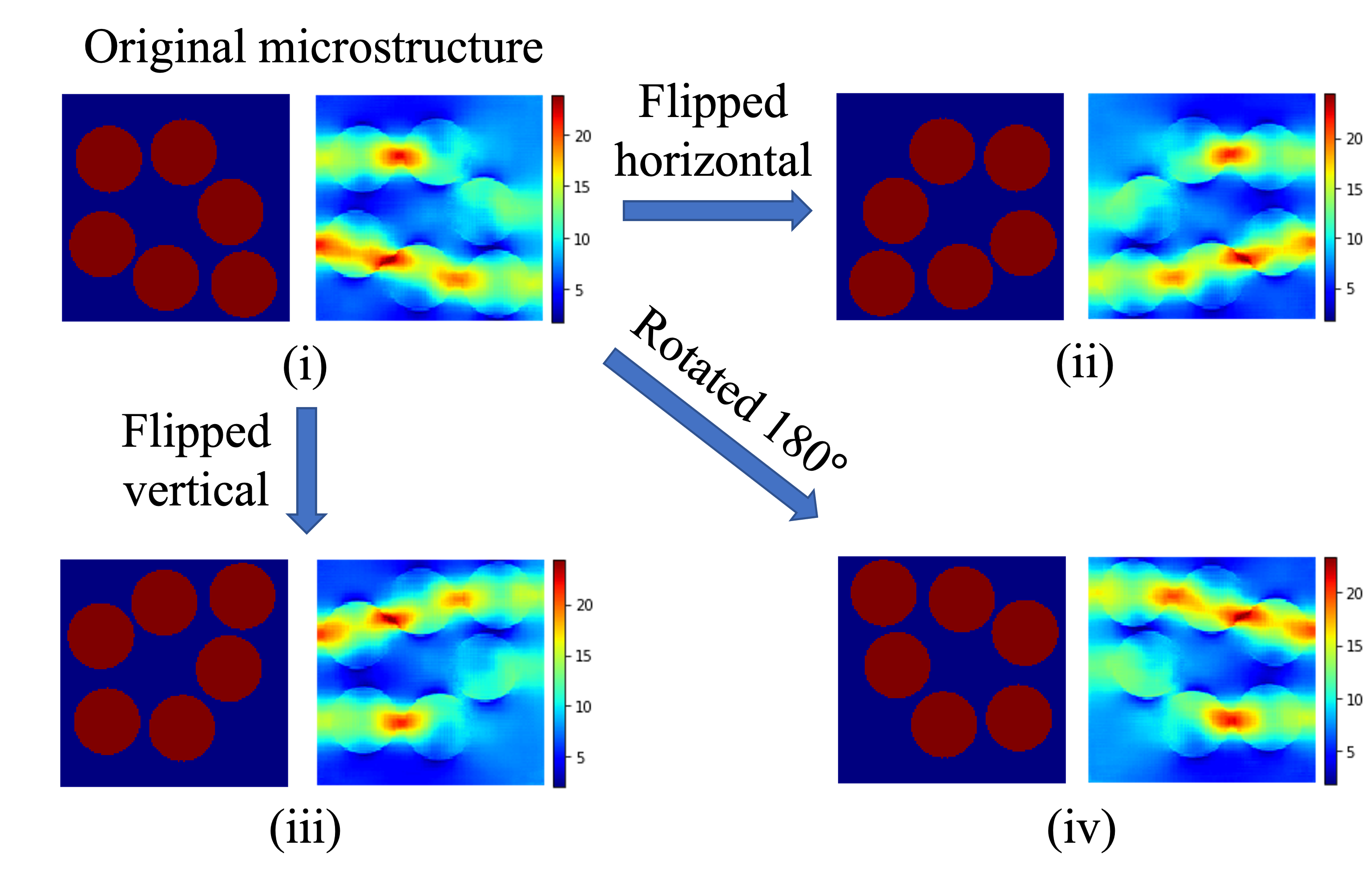}
    \includegraphics[width=0.48\columnwidth,keepaspectratio]{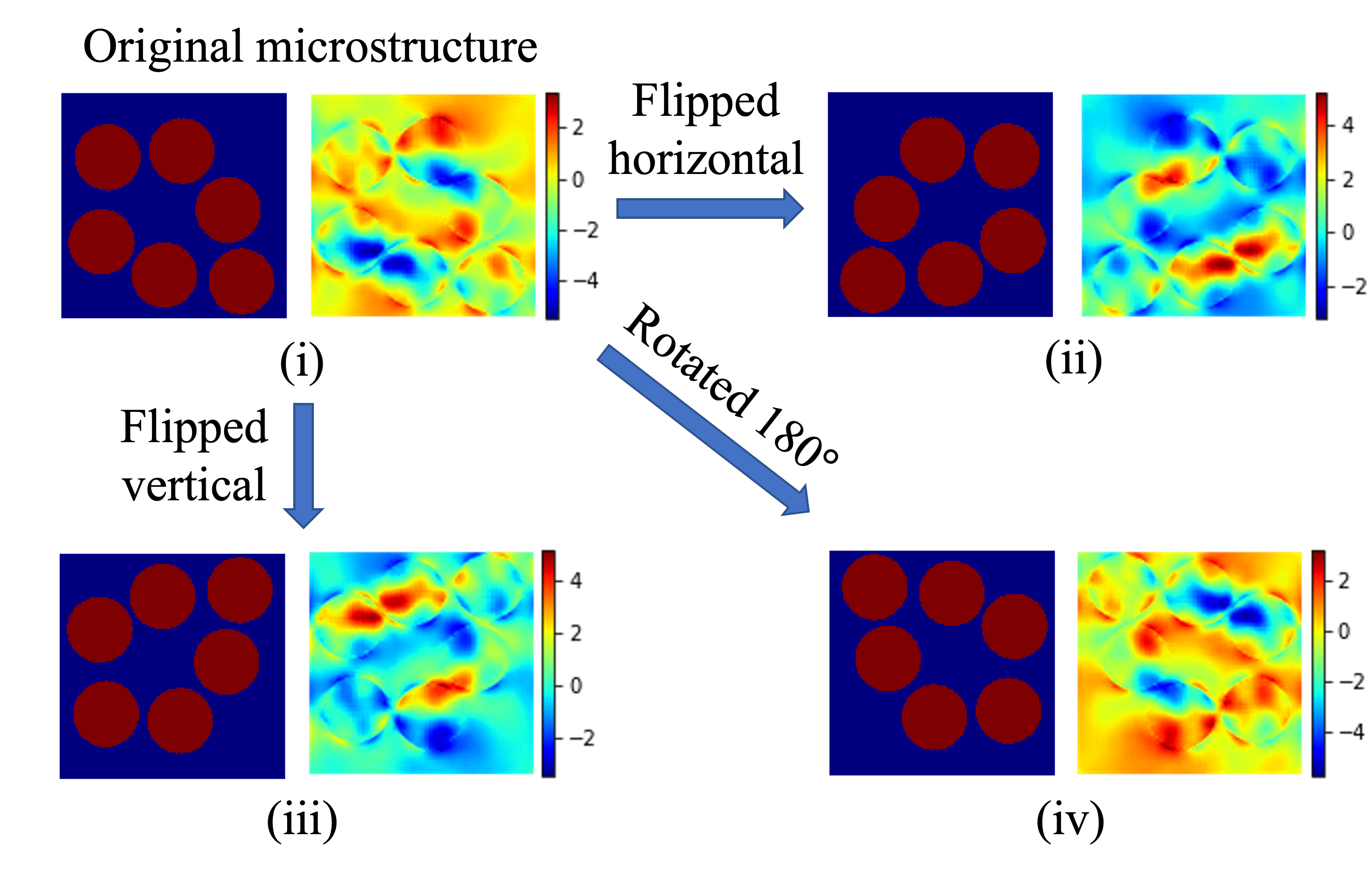}
    \hspace{2in} (a) \hspace{2in} (b)
    \caption{4-fold data augmentation by flipping (a) tensile and (b) shear stress maps associated with a tensile applied strain.}
    \label{dataAugmentation_tensileLoading}
\end{figure}

Data augmentation improves the robustness of the deep learning model, particularly in the case of mechanics problems in which there is often limited training data. Common data augmentation techniques in the image segmentation community include image flipping and rotation. A straightforward rotation of a stress map in a mechanics problem, however, would violate the underlying physics.  Therefore, a modified data augmentation approach consistent with the physics of the problem is implemented. The microstructure image is flipped sequentially  in the horizontal, vertical, and again horizontal plane. The stress maps corresponding to the flipped images are obtained by flipping them accordingly, as shown in Figure \ref{dataAugmentation_tensileLoading} for tensile and shear stress resulting from tensile strain loading. In this way, the training data is increased four-fold without performing any additional FE simulations.

\begin{figure}[h!]
    \centering
    \includegraphics[width=0.7\columnwidth,keepaspectratio]{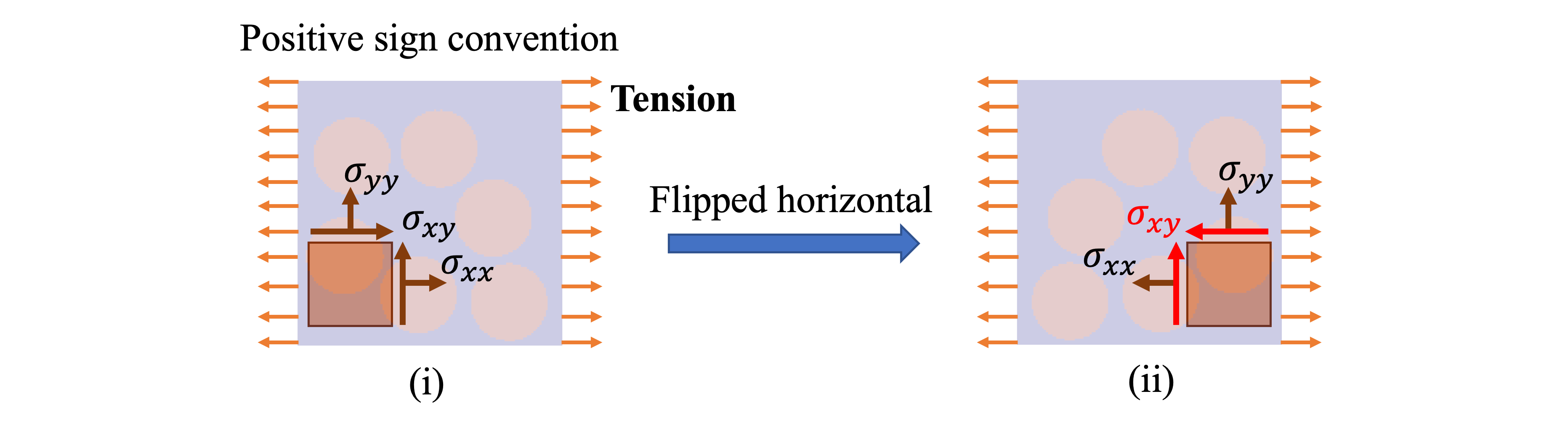}
    
    (a)
    
    \includegraphics[width=0.7\columnwidth,keepaspectratio]{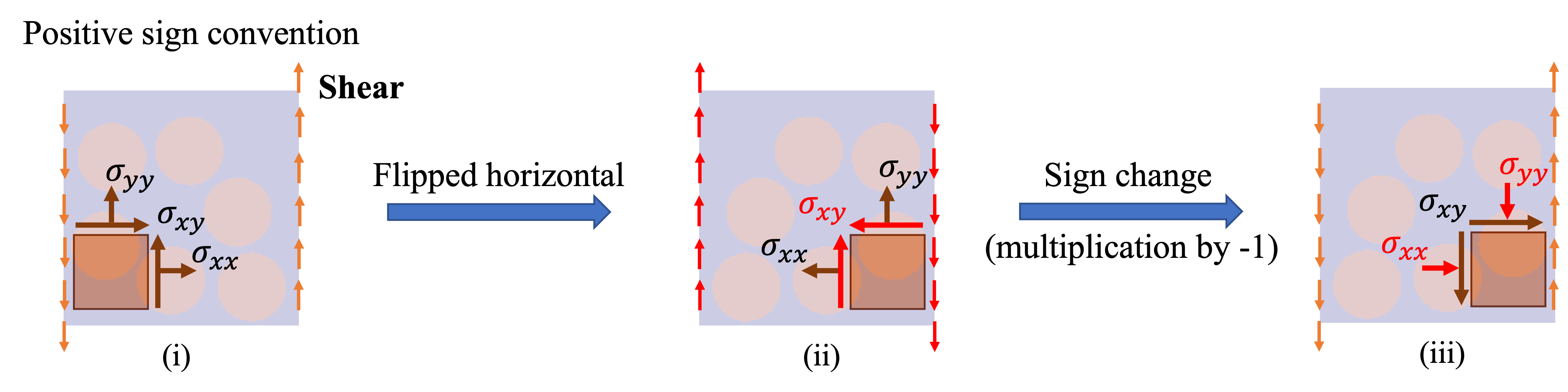}
   
    (b)
    \caption{Positive sign convention in case of (a) tensile and (b) shear loading. Under the applied tensile strain, the shear stress reverses direction when the image is flipped (shown in red). Under the applied shear strain, the shear boundary conditions reverse direction when the image is flipped, which is corrected by reversing the direction of all stress components, ultimately resulting in a change in direction of the normal stress (shown in red).}
    \label{dataAugmentation_signConvention_tensileLoading}
\end{figure}

Figure \ref{dataAugmentation_signConvention_tensileLoading}a shows that the tensile boundary conditions are symmetric, so they do not change direction when the image is flipped . However, the shear stress reverses direction under a horizontal flip, which is corrected by multiplying the shear stress by $-1$, as visualized in Figure \ref{dataAugmentation_tensileLoading}b. 

\begin{figure}[h!]
    \centering
    \includegraphics[width=0.48\columnwidth,keepaspectratio]{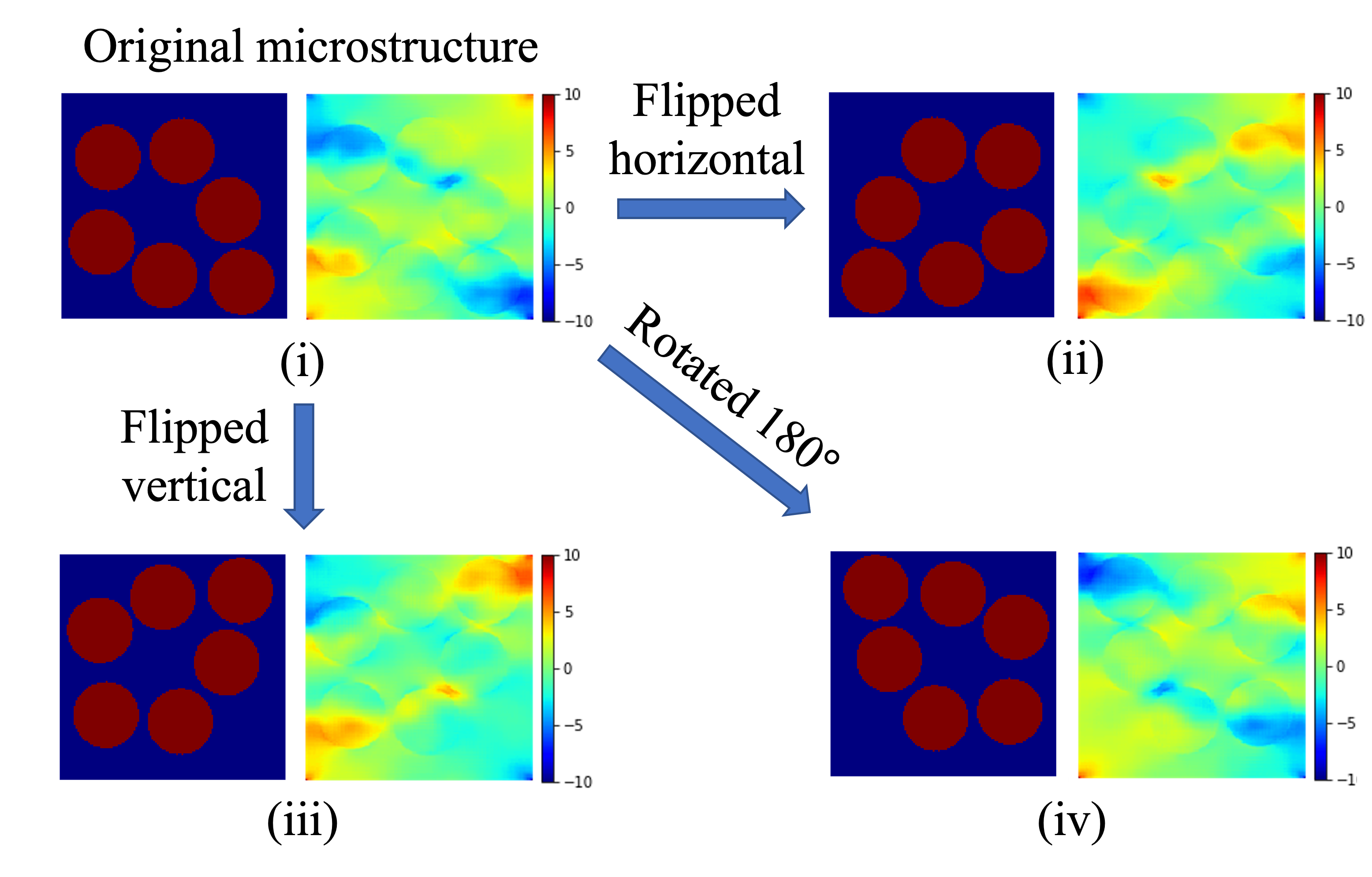}
    \includegraphics[width=0.48\columnwidth,keepaspectratio]{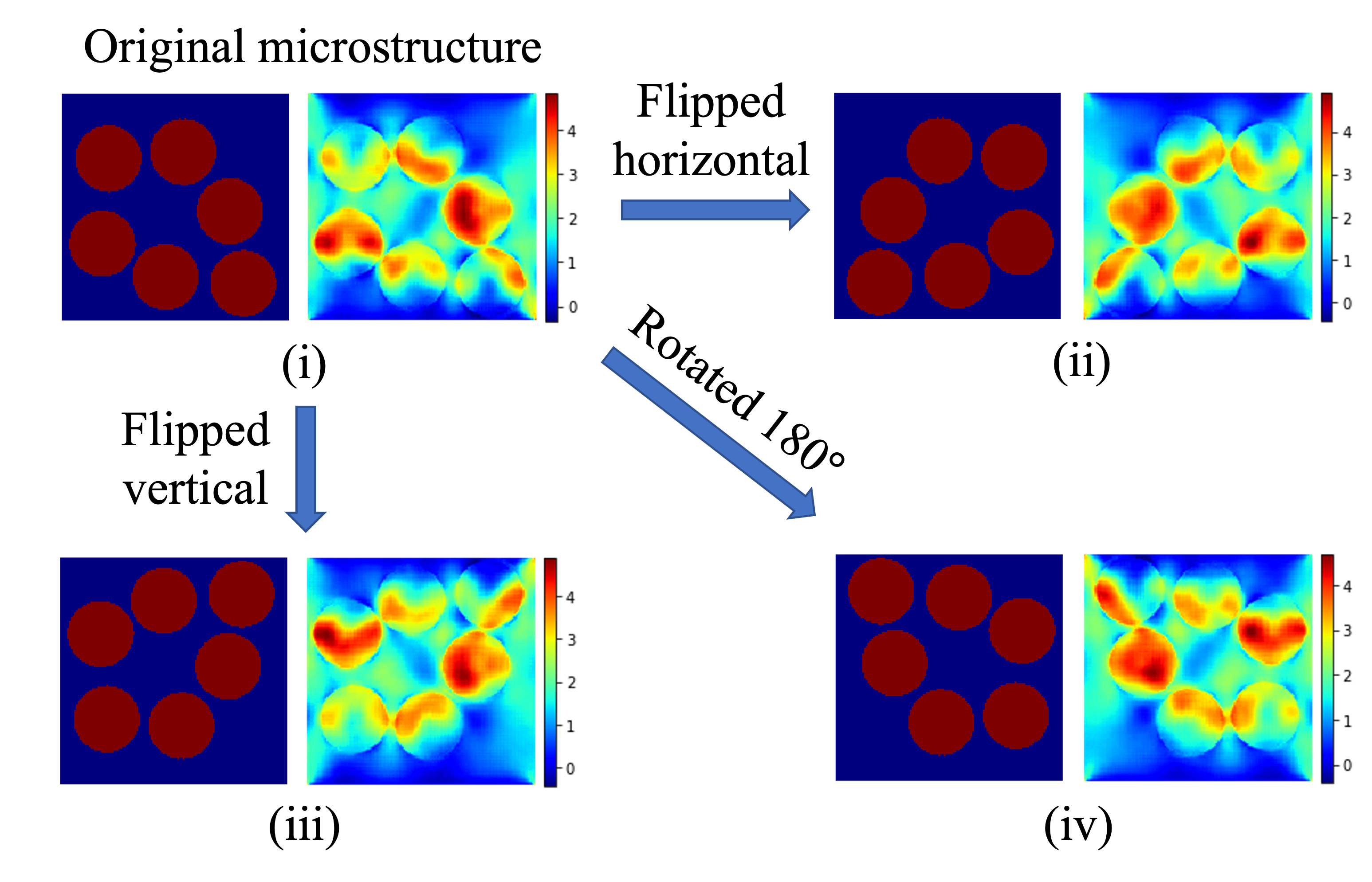}
    \hspace{2in} (a) \hspace{2in} (b)
    \caption{4-fold data augmentation by flipping (a) tensile and (b) shear stress maps associated with a shear applied strain.}
    \label{dataAugmentation_shearLoading}
\end{figure}

The shear boundary conditions are antisymmetric, so they reverse direction when the image is flipped (see Figure \ref{dataAugmentation_signConvention_tensileLoading}b). This is corrected by multiplying all stress components by $-1$, which causes the normal stresses to act in the opposite direction to the original direction and the shear stresses to revert to the original direction, as shown in Figure \ref{dataAugmentation_shearLoading}. These simple transformations shown in Figure \ref{dataAugmentation_tensileLoading} and \ref{dataAugmentation_shearLoading} result in four times the training data at negligible additional computational cost.

\subsection{Deep learning model} \label{sec:unet}
An encoder-decoder ML-model is trained to map the microstructure to the stress tensor field maps, avoiding the use of a physics based solver once fully trained. A single model is trained from scratch to predict the six components of the two 2D stress tensor fields resulting from the tensile and shear tests, as shown in Figure~\ref{image_stacking}. As a result, in this ML-model, the single-channel input contains only the microstructure image, and the six-channel output contains the six components of the stress tensor fields corresponding to the two defined displacement boundary conditions.

\indent The microstructure is discretized as a binary input array of size $128\times128$ elements, where $1$ represents the location of fiber, and $0$ represents the location of matrix. Similarly, the stress from the finer, irregular mesh in the FE model is interpolated to a grid data of size 128 $\times$ 128. Thus, both the input and the output images are of size 128 $\times$ 128 pixels. Because the stress components are of different average magnitudes, the stress maps are standardized before training, in order to improve model performance. Each stress component is shifted and scaled to mean $0$ and standard deviation $1$ before training. The standardization constants are calculated from the training data and kept constant for the test data. 

\begin{figure}[b!]
    \centering
    \includegraphics[width=0.7\columnwidth,keepaspectratio]{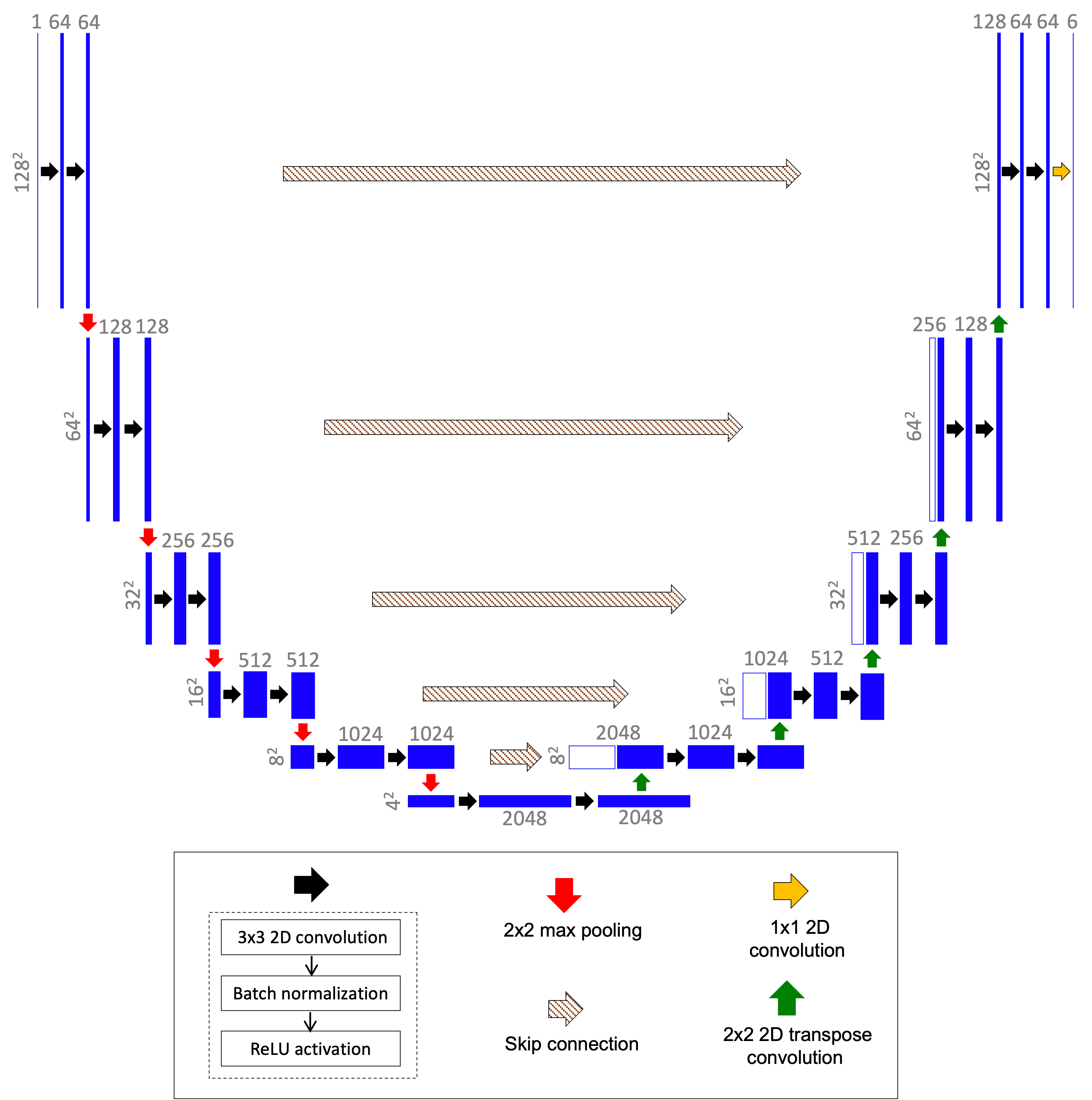}
    \caption{U-Net architecture with 4x4 pixels in the lowest resolution. Each blue box corresponds to a multi-channel feature map with the number of channels denoted on the top and channel height $\times$ width denoted on the left. White boxes represent concatenated feature maps. The arrows denote the different operations.}
    \label{UNet_architecture}
\end{figure}

\indent A U-Net deep learning architecture \cite{Ronneberger2015u} is used. The U-Net architecture consists of a contracting path that performs down-sampling and an expanding path that performs up-sampling. These two paths form a U-shaped network when connected together. The symmetric layers in the encoder-decoder architecture are connected using skip-connections that helps to recover fine-grained details in the prediction. This architecture is based on earlier work \cite{bhaduri2022stress} with a slight modification to generate a six-channel output as shown in Figure \ref{UNet_architecture}. The number of trainable parameters is approximately 124,000,000.

\indent The weighted mean squared error (MSE) between the true (FE simulated) and predicted (ML-model learned) stress tensor maps is used as the loss function: 

\begin{equation}
    \text{Mean weighted MSE} = \frac{1}{N_{\text{train}}}\sum_{k=1}^{N_{\text{train}}} \left[ \sum_{j=1}^{N_{\text{s}}} \left[ \frac{\sum_{i=1}^{S} |y_t^i| (y_p^i - y_t^i)^2}{\sum_{i=1}^{S} y_t^i} \right]_j \right]_k
\end{equation}

\noindent where $S = H \times W$, $H$ and $W$ are the height and width, respectively, of the image in pixels, $N_{\text{train}}$ is the number of training images, $N_{\text{s}}$ is the number of stress components, $y_t^i$ is the true stress value at pixel $i$ of a training image, and $y_p^i$ is the corresponding predicted stress value at the same pixel $i$. This weighting scheme improves prediction accuracy in high stress regions by assigning higher weights to the error associated with them. The deep learning model is built on TensorFlow \cite{abadi2016tensorflow} and run on Google Colaboratory Pro, a cloud-based Jupyter note-book environment that provides open access to a single NVIDIA Tesla P100 GPU with 32 GB memory. The ML-model achieves very good performance even when trained with just a few images. In particular, the ML-model is trained on $45$ data samples, which is augmented four times to $180$ data samples. The model parameters are fine-tuned on $20$ data samples in the validation dataset. The training is performed for $200$ epochs, which takes approximately $30$ minutes. Once trained, the model performance is evaluated on $1000$ data in the test dataset.

\subsubsection{Effect of training data size on prediction accuracy}\label{sec:train_size_effect}
\begin{figure}[t!]
    \centering
    \includegraphics[width=\columnwidth,keepaspectratio]{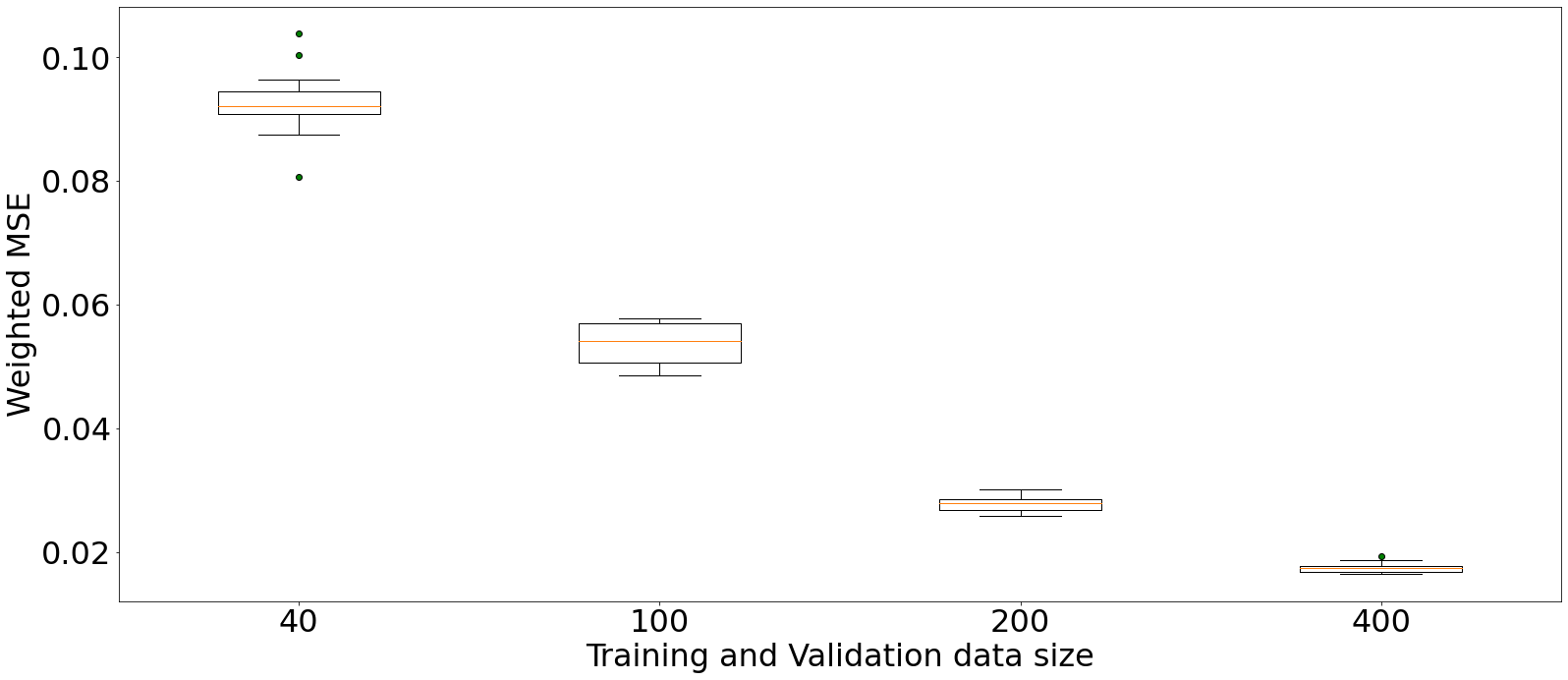}
    \caption{Effect of the number of images in the training \& validation data on the prediction accuracy of the ML-model.}
    \label{errConvPlot_20fib_trainingSize}
\end{figure}

The quality of the predicted stress maps depends on the number of images used to train the ML-model. To assess this, training is performed $20$ times with different random seed initializations of the ML-model parameters, and the accuracy of the ML-model is evaluated. Several cases are analyzed by varying the size of the training and validation data from $10$ to $100$, which is augmented four times to range from $40$ to $400$. We select the mean weighted MSE defined in Section \ref{sec:unet} as the accuracy metric. Figure \ref{errConvPlot_20fib_trainingSize} shows box plots of accuracy of the ML-model trained with different number of images. As expected, with increase in training data size, the mean accuracy increases, and the variance of the accuracy decreases. These results confirm that analyses based on $180$ training data samples provide reasonable accuracy.

\subsection{Effective material properties} \label{sec:effective_material_properties}
Subsequent analyses assume that the two-phase microstructure is linear elastic, with the elastic moduli $E_{fiber}$ and $E_{matrix}$ and Poisson's ratios $\nu_{fiber}$ and $\nu_{matrix}$, describing the fiber and matrix material, respectively. 
Even though the constituent phases are isotropic, the unidirectional, continuous fiber-reinforced composite will be orthotropic, with nine independent elastic constants \cite{herakovich1998mechanics}. In order to calculate these constants, the relationship between the macroscopic stress $\sigma_{ij}$ and strain $\varepsilon_{ij}$, in the principal material directions can be exploited: 

\begin{equation} \label{compliance_tensor}
\left\{\arraycolsep=1.2pt\def\arraystretch{1.5}
\begin{array}{l}
\varepsilon_{11} \\
\varepsilon_{22} \\
\varepsilon_{33} \\
2\varepsilon_{12} \\
2\varepsilon_{13} \\
2\varepsilon_{23}
\end{array}\right\}=\left[\arraycolsep=1.2pt\def\arraystretch{1.5}
\begin{array}{cccccc}
1 / E_{1} & -\nu_{21} / E_{2} & -\nu_{31} / E_{3} & 0 & 0 & 0 \\
-\nu_{12} / E_{1} & 1 / E_{2} & -\nu_{32} / E_{3} & 0 & 0 & 0 \\
-\nu_{13} / E_{1} & -\nu_{23} / E_{2} & 1 / E_{3} & 0 & 0 & 0 \\
0 & 0 & 0 & 1 / G_{12} & 0 & 0 \\
0 & 0 & 0 & 0 & 1 / G_{13} & 0 \\
0 & 0 & 0 & 0 & 0 & 1 / G_{23}
\end{array}\right]\left\{\arraycolsep=1.2pt\def\arraystretch{1.5}
\begin{array}{l}
\sigma_{11} \\
\sigma_{22} \\
\sigma_{33} \\
\sigma_{12} \\
\sigma_{13} \\
\sigma_{23}
\end{array}\right\}
\end{equation}

\noindent where the components of the full compliance matrix are expressed using the nine effective engineering properties of the equivalent homogeneous material: the three effective uniaxial moduli (Young’s moduli) $E_{1}$,$E_{2}$,$E_{3}$; effective Poisson's ratios $\nu_{12}$, $\nu_{13}$, $\nu_{23}$; and the effective shear moduli $G_{12}$, $G_{13}$, and $G_{23}$ associated with the principal material directions.

\indent The average strains and stresses from the plane strain FE analysis described in Section \ref{sec:methodology}, are used to calculate the effective material properties. Specifically, three mechanical tests are performed where the 2D microstructure is subjected to: (1) tensile strain loading along the x-direction, (2) tensile strain loading along the y-direction, and (3) in-plane shear strain loading. The average stresses are obtained by taking the average of the resulting stress map. The effective engineering constants are evaluated by substituting the average stresses and strains into Equation \ref{compliance_tensor}. The equations resulting from the two tensile loading tests are solved to evaluate the Young’s moduli $E_{1}$ and $E_{2}$ and effective Poisson's ratios $\nu_{12}$, $\nu_{13}$, and $\nu_{23}$. The in-plane shear modulus $G_{12}$ is evaluated from the shear loading condition. The Voigt approximation (rule of mixtures) has been demonstrated to be valid for estimating out of plane moduli  \cite{herakovich1998mechanics}: 

\begin{equation}
    E_{3} = E_{matrix} + (E_{fiber}-E_{matrix})V_{fiber}
\end{equation} 
\begin{equation}
    G_{13} = G_{matrix} + (G_{fiber}-G_{matrix})V_{fiber}
\end{equation} 
\begin{equation}
    G_{23} = G_{matrix} + (G_{fiber}-G_{matrix})V_{fiber}
\end{equation} 

\noindent where $V_{fiber}$ is the fraction of fiber in the composite.  An FE analysis with out-of-plane loading conditions can be carried out if more accurate estimates of the out of plane properties are desired, but the results do not change significantly enough to warrant the extra computational effort. Furthermore, the out-of-plane stresses do not have a significant effect on the in-plane behavior for the unidirectional composite material under study, so the simple and fast model of rule of mixtures proves to be sufficient for these purposes.

\subsection{ML-driven multiscale analysis} \label{sec:multiscale_formulation}
In the proposed ML-driven multiscale analysis approach, a two-scale modeling scheme is implemented by separately analyzing a heterogeneous macro-structure at both the macro and micro-scales as shown in Figure~\ref{overview}(c). The proposed approach improves on the $FE^2$ multiscale modeling scheme by using the efficient ML-model at the micro-scale and performing computationally expensive FE analysis only at the macro-scale. The macro-scale model is built by performing FE discretizations of the macro-structure using the effective elastic properties of the associated microstructure in each element, calculated by performing upscaling (homogenization) using the ML-approach, as outlined in Section~\ref{sec:effective_material_properties}. This macro-scale FE model provides the average stress/strain in every element. Applying these average strain values to the microstructure underlying that element, the ML-model provides local stress/strain distributions in the microstructure. Section~\ref{sec:results} demonstrates the accuracy of the ML-driven multiscale analysis in performing both homogenization and localization operations. The numerical examples demonstrate that the proposed ML-driven multiscale analysis is applicable to a variety of macro-structure sizes, loadings, and boundary conditions. Furthermore, the trained ML-model is applicable to both upscaling (homogenization) and downscaling (localization), and therefore requires only the initial training process, making the entire training process and the associated data requirements highly efficient. 

\section{Results} \label{sec:results}
This section describes the performance of the ML-driven multiscale analysis approach. 
Section \ref{sec:MicroscaleStressTensor} shows the accuracy of the ML-model to predict the micro-scale stress tensor. Section \ref{sec:pixel_physics} shows that the predicted stress maps satisfy the relevant physics at all points within the microstructure. Section \ref{sec:avg_stress_pred} shows that the ML-model also predicts the average stresses with high accuracy.
Multiscale analysis is achieved by using the pretrained ML-model to perform efficient homogenization and localization, which is shown in Sections \ref{sec:homogenization_U_Net} and \ref{sec:mult_load}, respectively. Section \ref{sec:mult_u_net} demonstrates the efficacy of the ML-driven approach in performing multiscale mechanics modeling, through several numerical examples. A validation of the predicted results with full-scale FE analyses confirms that the proposed approach performs at a remarkable speed with high accuracy.

\begin{figure}[b!]
    \centering
    \includegraphics[width=0.7\columnwidth,keepaspectratio]{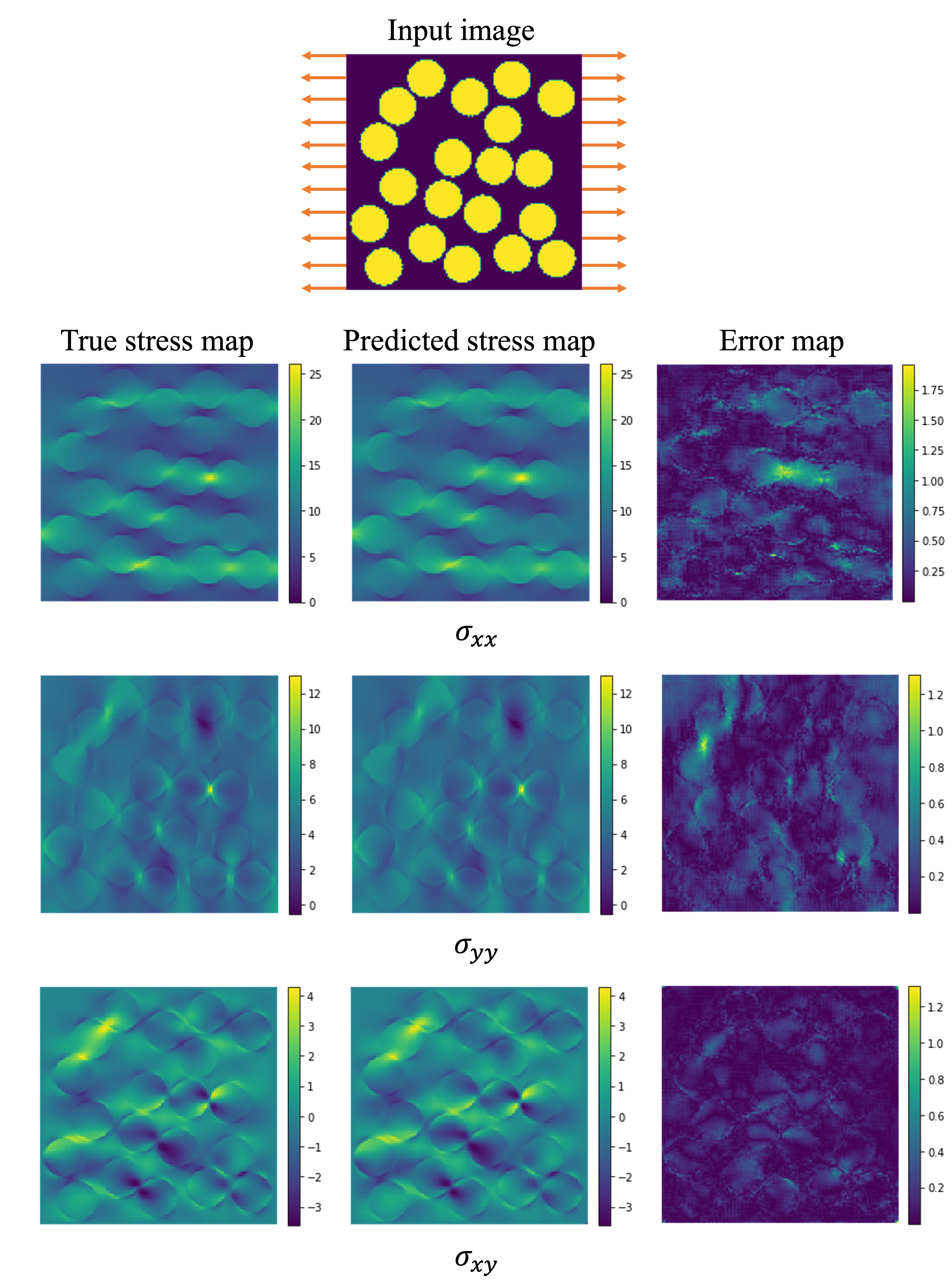}
    \caption{Stress tensor map resulting from 0.1\% tensile strain loading predicted from the ML-model.}
    \label{prediction_N200_tensileLoading}
\end{figure}

\begin{figure}[b!]
    \centering
    \includegraphics[width=0.7\columnwidth,keepaspectratio]{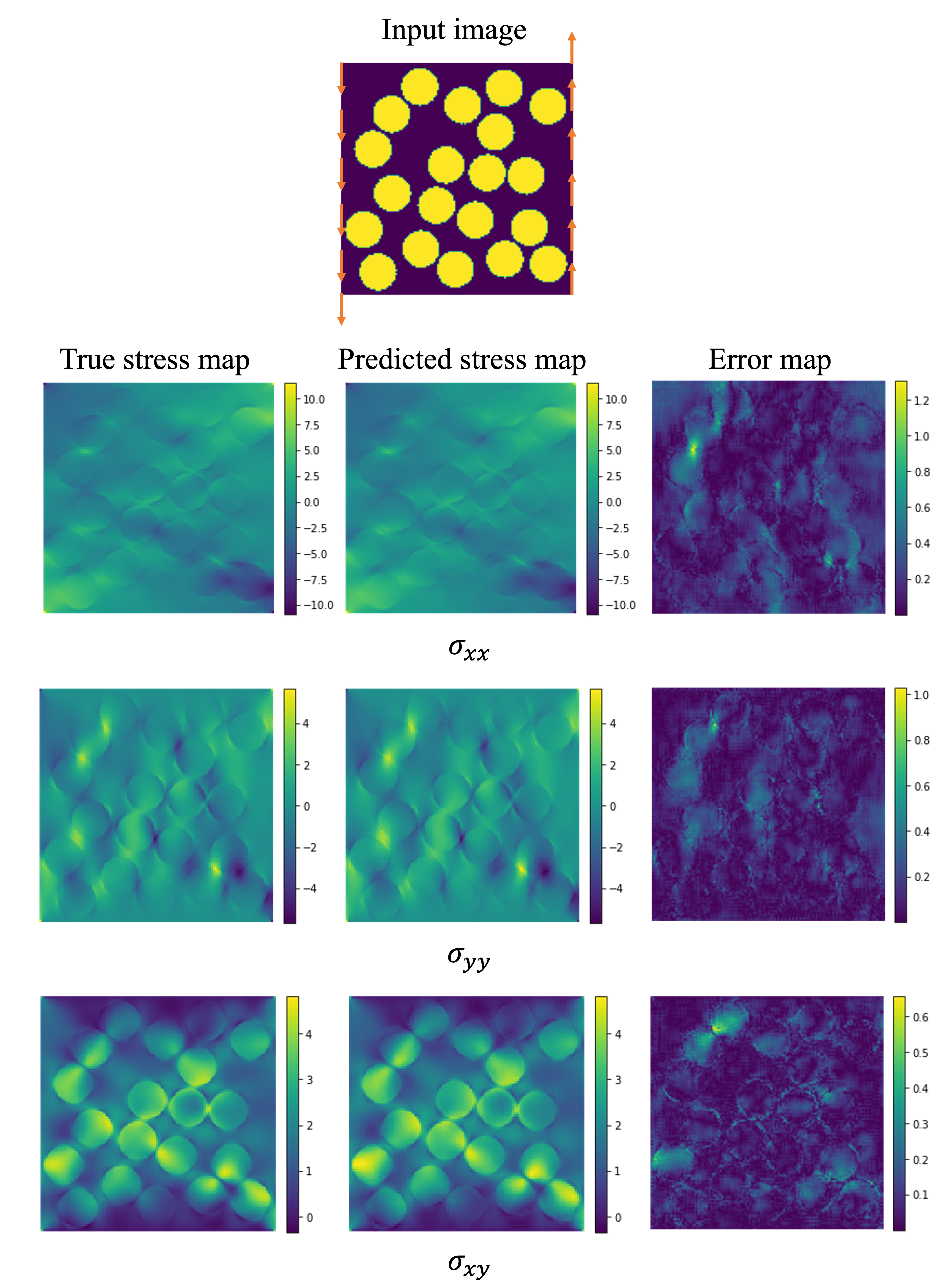}
    \caption{Stress tensor maps resulting from 0.1\% shear strain loading predicted from the ML-model.}
    \label{prediction_N200_shearLoading}
\end{figure}

\subsection{Micro-scale stress tensor prediction accuracy}\label{sec:MicroscaleStressTensor}
The ML-model directly maps the composite microstructure to the corresponding stress maps resulting from a $0.1\%$ tensile and $0.1\%$ shear strain loading. Figure \ref{prediction_N200_tensileLoading} shows the prediction results for a sample microstructure image from the test dataset subjected to tensile strain loading. Figure \ref{prediction_N200_tensileLoading} shows the true (FE simulated) and predicted (ML-model learned) stress maps. The pixel-wise absolute error between the predicted and true stress maps is used as the error metric, defined as:
\begin{equation}
    \text{Absolute error} = \left| {\sigma_{p}^i-\sigma_{t}^i} \right|
\end{equation}

\noindent where $\sigma_t^i$  is the true stress value at pixel $i$ of a test image and $\sigma_p^i$ is the corresponding predicted stress value at the same pixel $i$. The absolute error maps in Figure \ref{prediction_N200_tensileLoading} indicate that the prediction error is relatively small. 
Figure \ref{prediction_N200_shearLoading} shows similar plots for the stress tensor maps resulting from shear strain loading. In this way, the ML-model provides a total of six stress images for each microstructure image. These plots show that the ML-model predicts the micro-scale stress maps resulting from a $0.1\%$ tensile and $0.1\%$ shear loading with good accuracy. The ML-model requires ~0.1 seconds to analyze a microstructure, which is more than 600 times faster than FE analysis on the same computational platform.

\subsection{Pixel-wise physics validation}\label{sec:pixel_physics}
\begin{figure}[h!]
    \centering
    \includegraphics[width=0.7\columnwidth,keepaspectratio]{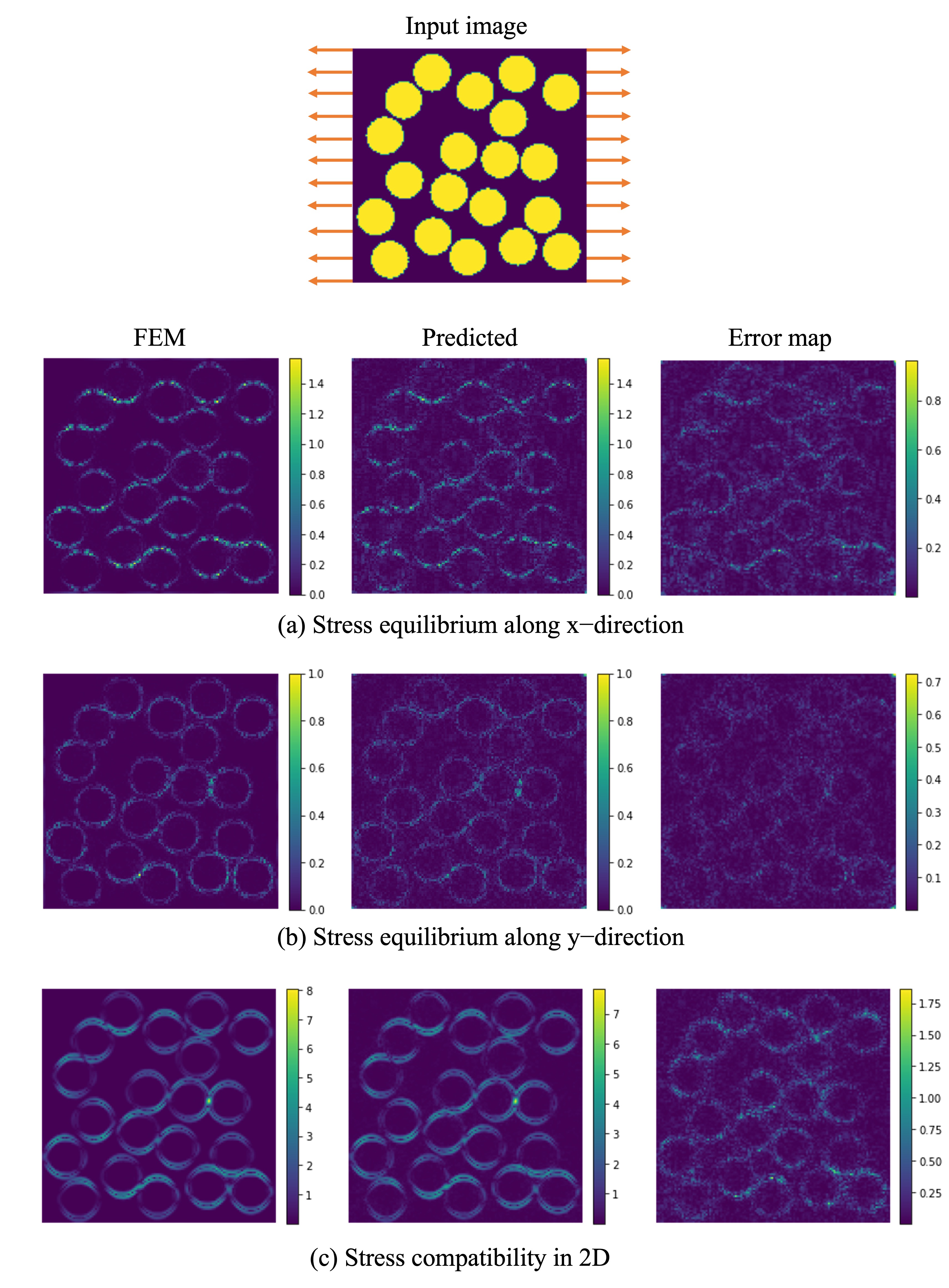}
    \caption{Pixel-wise physics validation for the FE simulated and predicted stress maps by evaluating: (a) equilibrium equation along the x-direction, (b) equilibrium equation along the y-direction, and (c) stress compatibility equation at each pixel location in the microstructure.}
    \label{physics_equations_tensile}
\end{figure}

Another verification of the ML-model is to confirm that the predicted stress maps also satisfy the relevant physics. To verify this, the field equations of elasticity theory \cite{sadd2009elasticity} are evaluated at each pixel location in the microstructure. In particular, the 2D stress equilibrium equations,

\begin{equation}
    \sigma_{{xx},x} + \sigma_{{xy},y}  = 0
\end{equation}
\begin{equation}
    \sigma_{{yy},y} + \sigma_{{xy},x}  = 0,
\end{equation}

\noindent and the Beltrami-Michell stress compatibility equation,

\begin{equation}
    \sigma_{{xx},xx} + \sigma_{{xx},yy} + \sigma_{{yy},xx} + \sigma_{{yy},yy}  = 0,
\end{equation}

\noindent are evaluated at each pixel using a finite difference scheme. Figures \ref{physics_equations_tensile}(a) and \ref{physics_equations_tensile}(b) show that the stress equilibrium conditions are satisfied consistently with the finite element results, and Figure \ref{physics_equations_tensile}(c) shows that the compatibility condition is satisfied at all pixels consistently with the finite element results. Figure \ref{physics_equations_tensile} shows that the small lack of equilibrium and compatibility at the interfaces is consistent between FE analysis and the ML-model predictions. This error at the fiber-matrix interfaces is not a specific shortcoming of the ML-model, but it is instead associated with the element discretization in the underlying FE models.

\subsection{Average stress prediction accuracy}\label{sec:avg_stress_pred}
\begin{figure}[h!]
    \centering
    \includegraphics[width=\columnwidth,keepaspectratio]{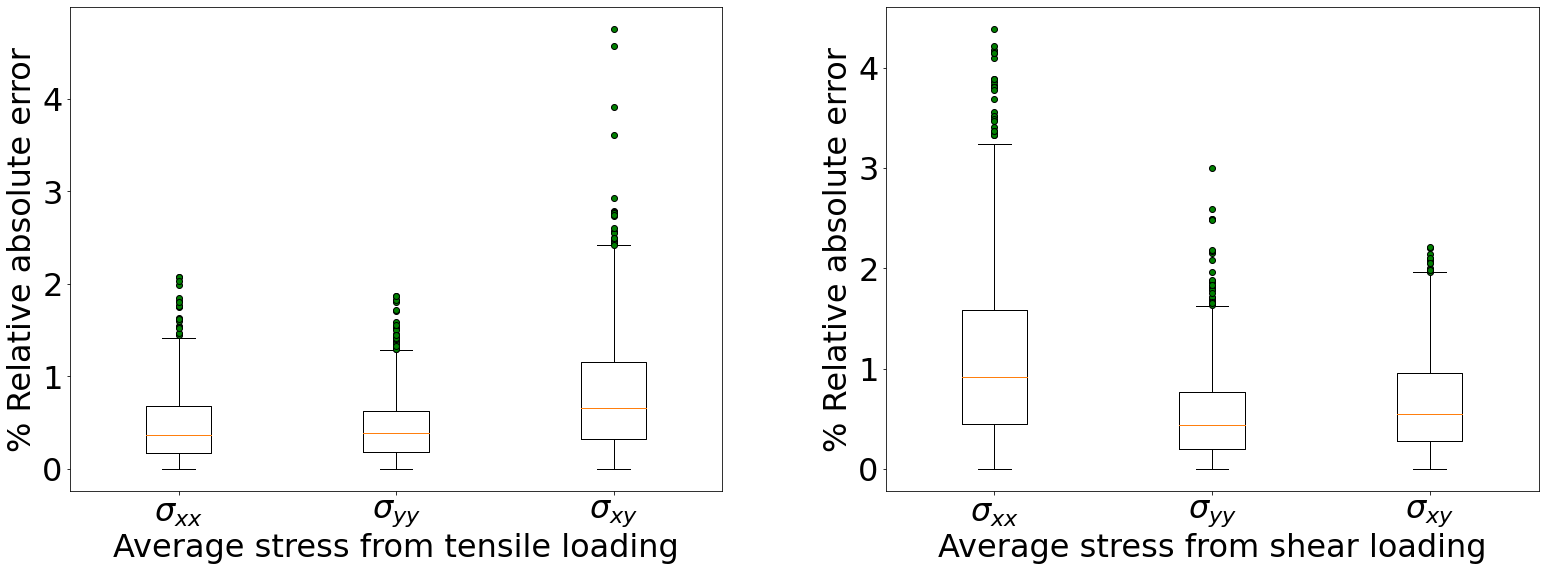}
    \caption{Error in average stresses prediction by using the ML-model as compared to FE analysis resulting from (a) a $0.1\%$ tensile strain loading, and (b) a $0.1\%$ shear strain loading.}
    \label{box_plot_avg_stress_pred}
\end{figure}

The spatial average of the local stress fields predicted by the ML-model are used to predict the effective properties during the homogenization step of the multiscale analysis. In order to assess the accuracy of this model for homogenization purposes, the relative absolute error between the predicted (ML-model learned) and true (FE simulated) average stress is calculated:
\begin{equation}
    \text{Relative absolute error} = \left| \frac{\bar{\sigma}_{p}-\bar{\sigma}_{t}}{\max\limits_{i}({\sigma}_{t}^i)} \right|
\end{equation}

\noindent where $\bar{\sigma}_t$ is the true average stress value and $\bar{\sigma}_p$ is the corresponding predicted average stress value. Here, $\max\limits_{i}({\sigma}_{t}^i)$ is the maximum value of the true stress over all the pixels in the stress map. The maximum stress value is preferred for this normalization, because normalizing by the true stress is problematic if the true stress value is $0$. Figure \ref{box_plot_avg_stress_pred} shows box plots of the relative absolute error between the true and predicted average stresses for the $1000$ test images. 
These plots show that the ML-model predicts the average stresses with high accuracy. 

\subsection{Homogenization using the ML-model}\label{sec:homogenization_U_Net}
\begin{figure}[h!]
    \centering
    \includegraphics[width=\columnwidth,keepaspectratio]{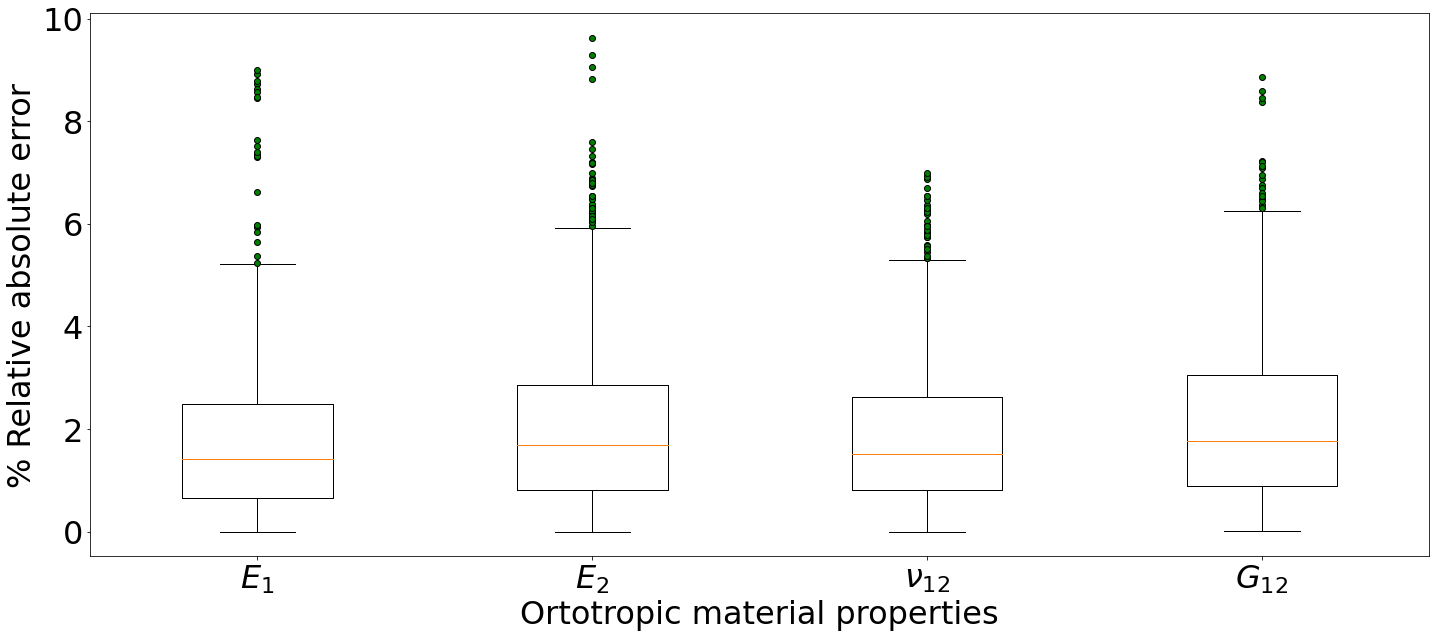}
    \caption{Error in the effective material properties prediction by using the ML-model as compared to FE analysis}
    \label{box_plot_avg_orthotropicProp}
\end{figure}

The average stresses predicted from the ML-model in Section \ref{sec:avg_stress_pred} and the corresponding applied strains are used to calculate the effective material properties based on the approach described in Section \ref{sec:effective_material_properties}. The predicted compliance matrix satisfies the theoretical upper and lower bounds as defined by Voigt and Reuss approximation of rule of mixtures \cite{herakovich1998mechanics}. The relative absolute error between the predicted (ML-model learned) and true (FE simulated) effective material properties is used as the error metric:
\begin{equation}
    \text{Relative absolute error} = \left| \frac{\bar{c}_{p}-\bar{c}_{t}}{\bar{c}_{t}} \right|
\end{equation}

\noindent where $\bar{c}_t$ is the true effective property value and $\bar{c}_p$ is the corresponding predicted effective property value. Figure \ref{box_plot_avg_orthotropicProp} shows box plots of the relative absolute error between the true and predicted in-plane elastic constants: $E_{1}$, $E_{2}$, $\nu_{12}$, and $G_{12}$ for the $1000$ test images. These plots show that the ML-model predicts the effective material properties with high accuracy. In this way, the ML-model performs efficient homogenization in a multiscale analysis framework.

\subsection{Multiaxial loading using the ML-model}\label{sec:mult_load}
The ML-model can predict local stresses under a multiaxial strain state, by combining the uniaxial tensile strain and in-plane shear strain loading through the principle of superposition. The superposition principle for a linear elastic material states that the resulting stress from a combination of loads is the sum of the stresses from each independent load applied to the structure \cite{sadd2009elasticity}. For a microstructure subjected to $\alpha\%$ tensile strain along the x-direction, $\beta\%$ tensile strain along the y-direction, and $\gamma\%$ in-plane shear strain loading, the resulting tensile stress along x-direction in the microstructure is calculated as:

\begin{equation}
     \sigma_{xx}^{(total)} = \frac{\alpha}{0.1} \sigma_{xx}^{(1)} + \frac{\beta}{0.1} \sigma_{xx}^{(2)} + \frac{\gamma}{0.1} \sigma_{xx}^{(3)}
\end{equation}

\noindent where $\sigma_{xx}^{(1)}$ and $\sigma_{xx}^{(2)}$ are the tensile stress resulting from a $0.1\%$ tensile strain loading along the x- and y-direction, respectively, $\sigma_{xx}^{(3)}$ is the tensile stress resulting from a $0.1\%$ in-plane shear strain loading, and $\sigma_{xx}^{(total)}$ is the resulting tensile stress from the combined multiaxial strain loading. A similar calculation can be performed to calculate the other stress components.  

\begin{figure}[h!]
    \centering
    \includegraphics[width=0.7\columnwidth,keepaspectratio]{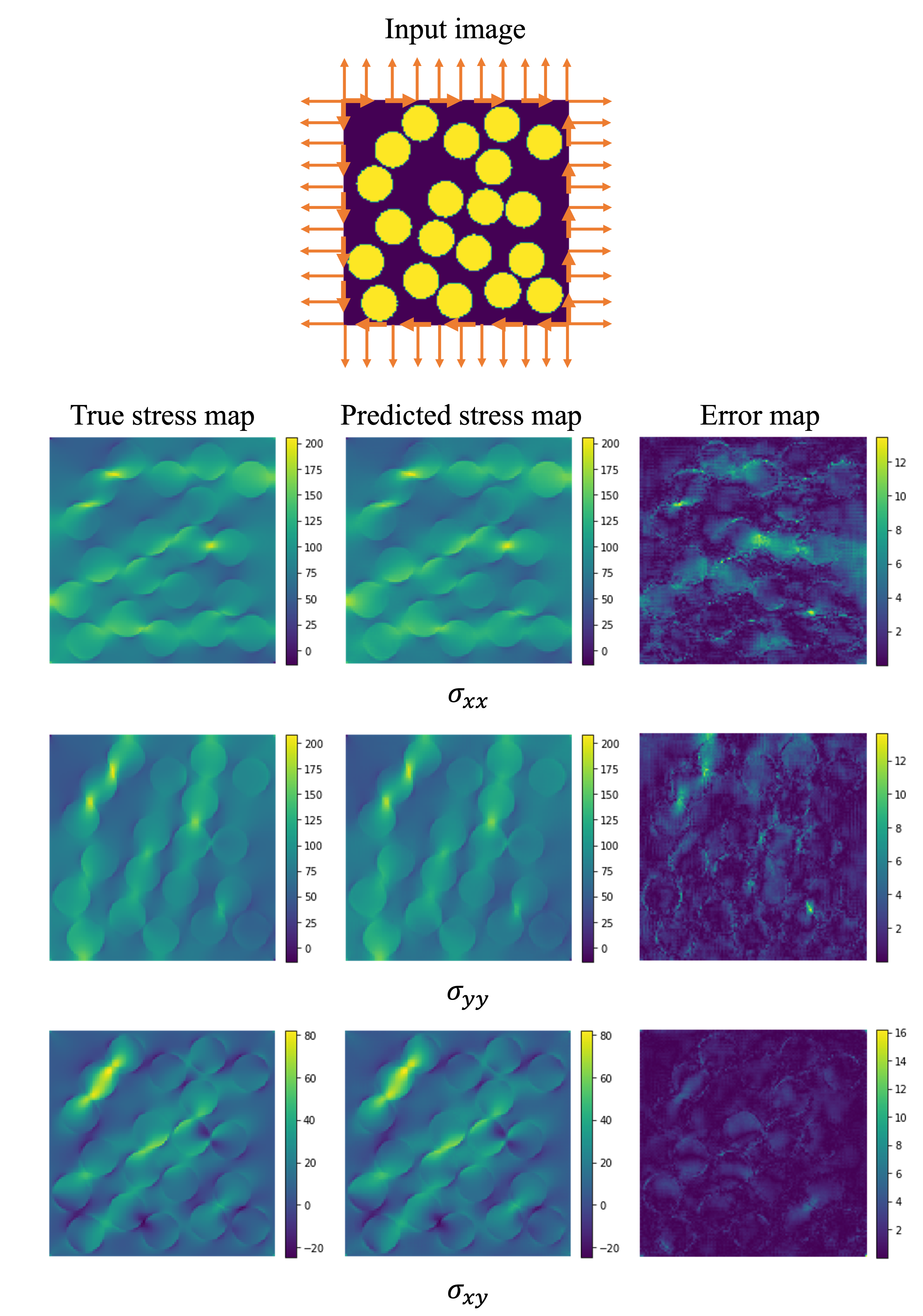}
    \caption{Stress tensor map resulting from a multiaxial loading predicted from the ML-model.}
    \label{prediction_N200_multiAxialLoading}
\end{figure}

\indent The ML-model predicts the stress maps for uniaxial tensile strain along the x-direction and in-plane shear strain loading as already described in Section \ref{sec:MicroscaleStressTensor}. The same model can be used to predict the stress maps for uniaxial tensile strain loading along the y-direction by a simple $90^{\circ}$ rotation. Consequently, the solution to a more complicated multiaxial strain loading case is predicted by scaling and adding the results from the ML-model for uniaxial tensile strain along the x- and y-direction, and in-plane shear strain loading. Figure \ref{prediction_N200_multiAxialLoading} shows the true (FE simulated) and predicted (ML-model learned) stress maps resulting from the microstructure subjected to $0.7\%$ tensile strain along the x-direction, $0.5\%$ tensile strain along the y-direction, and $0.8\%$ shear strain loading. These plots show that the ML-model accurately predicts the micro-scale stress maps resulting from multiaxial loading. In this way, the ML-model performs efficient localization in a multiscale analysis framework.

\subsection{ML-driven multiscale analysis}\label{sec:mult_u_net}
Numerical examples are provided in this section to demonstrate the effectiveness of the ML-driven multiscale analysis approach based on a pretrained ML-model. The pretrained ML-model, trained previously for the task of stress prediction in the microstructure, is used to perform both upscaling (homogenization) and downscaling (localization) to support multiscale analysis. Three macro-structures of varying sizes and subjected to different loading and boundary conditions are studied. 
 These numerical examples demonstrate that the proposed ML-driven multiscale analysis is applicable to a variety of macro-structure sizes, loadings, and boundary conditions. Furthermore, the ML-driven approach performs multiscale analysis at a remarkable speed with good accuracy when compared to conventional multiscale FE approaches, such as $FE^2$.

\subsubsection{Stress prediction in a $4\times$\ larger geometry}
\begin{figure}[h!]
    \centering
    \includegraphics[width=0.7\columnwidth,keepaspectratio]{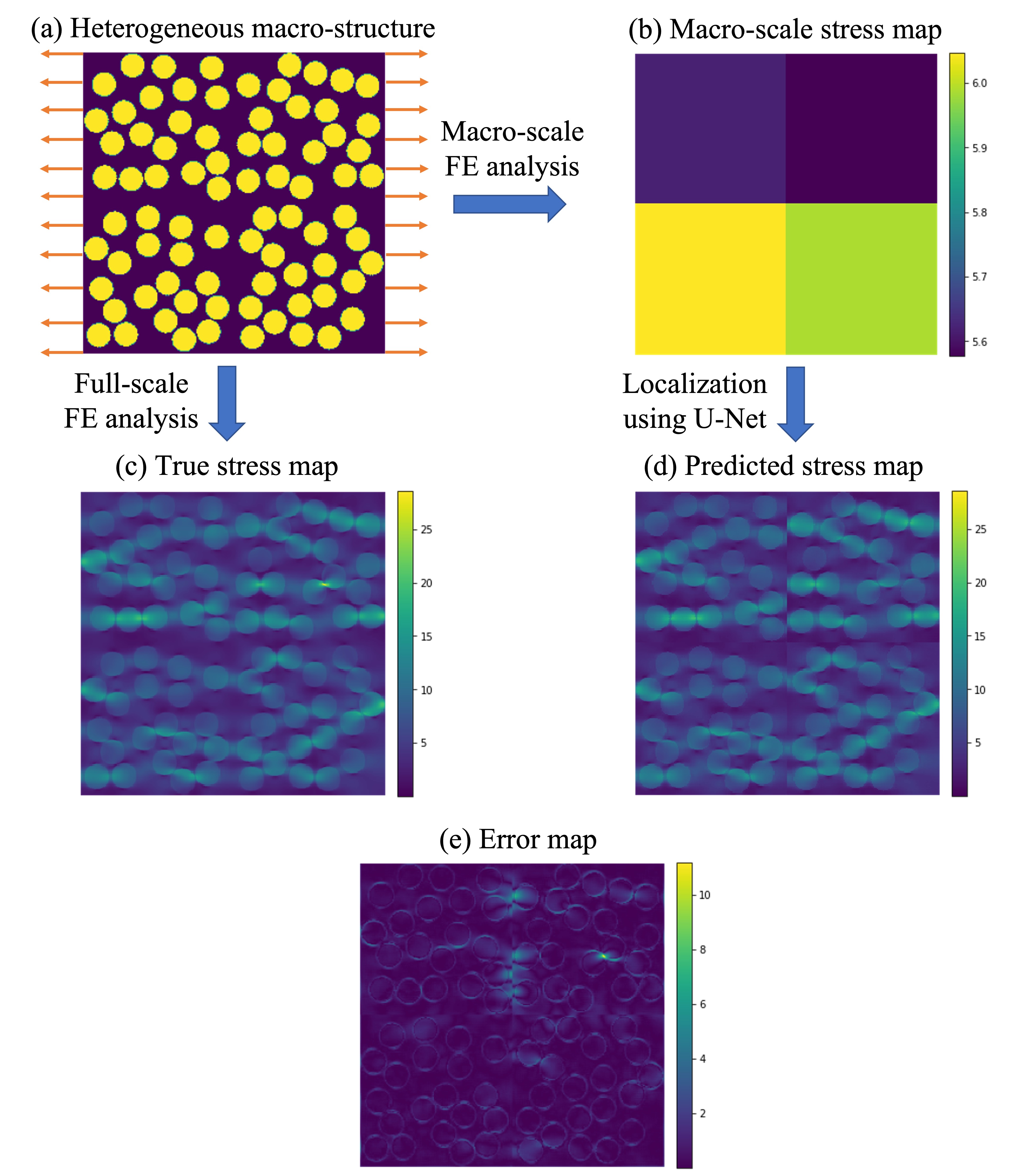}
    \caption{Multiscale analysis of a square domain of size $4\times$\ larger than the original microstructure subjected to $0.1\%$ tensile strain loading along the x-direction. (a) The boundary condition applied to the heterogeneous macro-structure. The von Mises stress maps resulting from: (b) macro-scale FE analysis of the $2\times2$ discretized square domain, (c) the subsequent ML-driven multiscale analysis, and (d) the full-scale FE analysis. (e) The pixel-wise absolute error between the FE simulated and predicted stress maps.}
    \label{multiscale_vonMises_2x2}
\end{figure}

The ML-model is used to perform a multiscale analysis of a square domain of size $4\times$\ larger than the original microstructure as shown in Figure \ref{multiscale_vonMises_2x2}(a). The square domain is subjected to a $0.1\%$ tensile strain loading along the x-direction as illustrated in Figure \ref{multiscale_vonMises_2x2}(a). This square domain is discretized by $2\times2$ square FE elements. The effective macroscopic properties of each element are calculated by performing homogenization using the ML-model. This $2\times2$ square domain is then analyzed by performing a macro-scale plane strain FE analysis to get the resulting average stress/strain in each element. Figure \ref{multiscale_vonMises_2x2}(b) shows the average von Mises stress in this $2\times2$ square domain. 
Localization is performed by using the ML-model to predict the local stress response, based on the average strain state in each of the $4$ elements in the $2\times2$ square domain. 
In this way, the ML-driven multiscale analysis approach predicts the resulting 2D stress tensor fields; namely, $\sigma_{xx}$, $\sigma_{yy}$ and $\sigma_{xy}$ at every microstructural location in the macrostructure. These are used to calculate von Mises stress, $\sigma_{v}$:

\begin{equation}
    \sigma_{v} = \sqrt{\frac{1}{2}\left[ (\sigma_{xx}-\sigma_{yy})^2 + (\sigma_{yy}-\sigma_{zz})^2 + (\sigma_{zz}-\sigma_{xx})^2 + 3(\sigma_{xy}^2) \right]}
\end{equation}

\noindent where $\sigma_{zz} = \left[ \nu_{fiber} \ I_{(fiber)} + \nu_{matrix} \ I_{(matrix)} \right] (\sigma_{xx} + \sigma_{yy})$ is implicit in the plane strain assumption, with $I_{()}$ being the binary indicator function. Figure \ref{multiscale_vonMises_2x2}(c) and \ref{multiscale_vonMises_2x2}(d) show the von Mises stress maps resulting from the ML-driven multiscale analysis and full-scale FE analysis, respectively. The absolute error map in Figure \ref{multiscale_vonMises_2x2}(e) indicates that the pixel-wise absolute error between the FE simulated and predicted stress maps is relatively small. Although there is a small error at the element interfaces, this is not a specific shortcoming of the ML-model, but it is instead associated with the domain discretization in the underlying two-scale approach. Similar error at the element boundaries would arise using $FE^2$ or other similar approaches.

\subsubsection{Stress prediction in a $16\times$\ larger geometry}
\begin{figure}[h!]
    \centering
    \includegraphics[width=0.7\columnwidth,keepaspectratio]{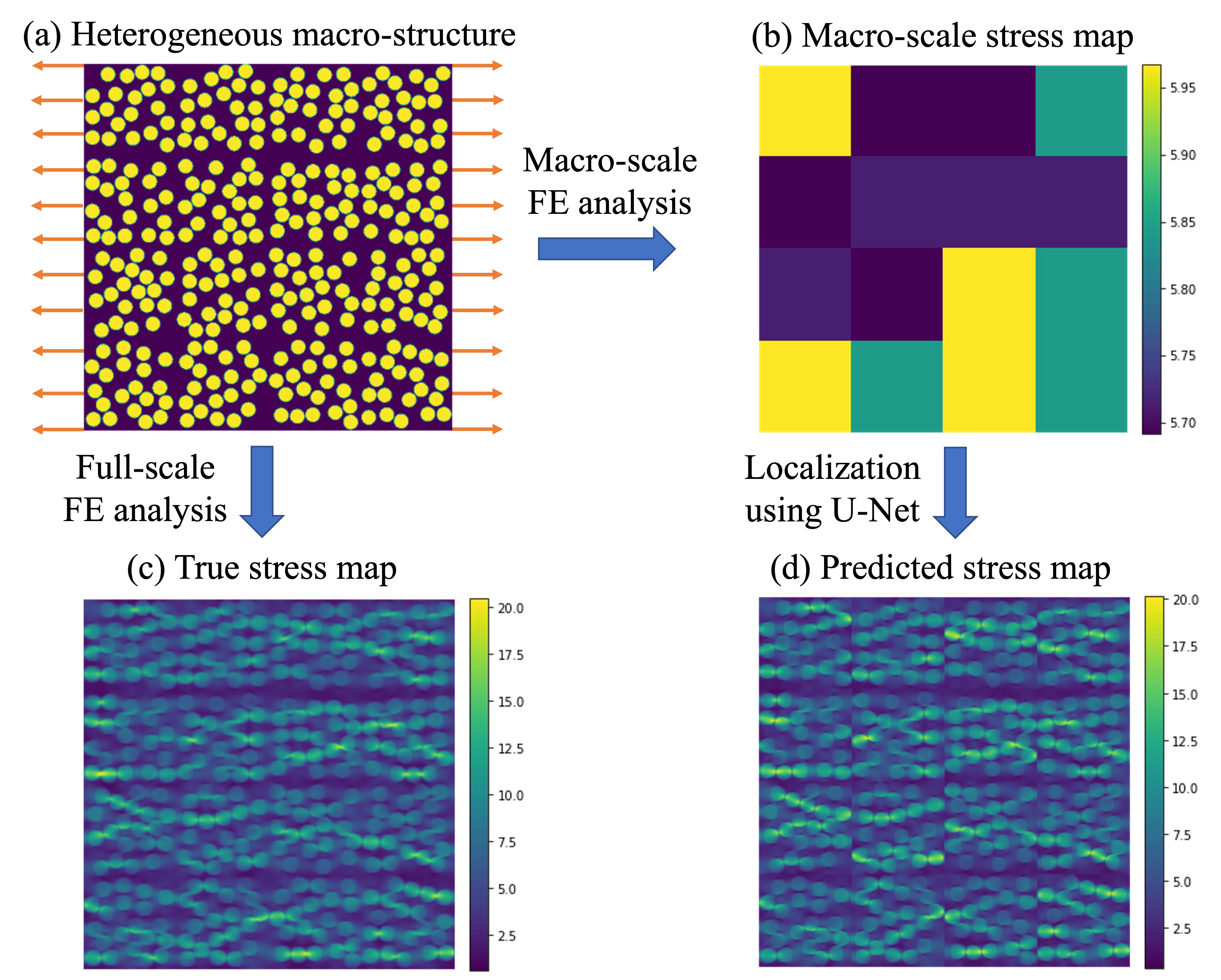}
    \caption{Multiscale analysis of a square domain of size $16\times$\ larger than the original microstructure subjected to $0.1\%$ tensile strain loading along the x-direction. (a) The boundary condition applied to the heterogeneous macro-structure. The von Mises stress maps resulting from: (b) macro-scale FE analysis of the $4\times4$ discretized square domain, (c) the subsequent ML-driven multiscale analysis, and (d) the full-scale FE analysis.}
    \label{multiscale_vonMises_4x4}
\end{figure}

A multiscale analysis of a square domain of size $16\times$\ larger than the original microstructure is performed as shown in Figure \ref{multiscale_vonMises_4x4}(a). The square domain is subjected to a $0.1\%$ tensile strain loading along the x-direction as illustrated on Figure \ref{multiscale_vonMises_4x4}(a). This square domain is discretized by $4\times4$ square FE elements. Figure \ref{multiscale_vonMises_4x4}(b) shows the average von Mises stress in this $4\times4$ square domain from the macro-scale plane strain FE analysis. Figure \ref{multiscale_vonMises_4x4}(c) and \ref{multiscale_vonMises_4x4}(d) show the predicted von Mises stress maps resulting from the full-scale FE analysis and the ML-driven multiscale analysis, respectively. These plots show that the ML-model predicts the micro-scale stress map with good accuracy, where any errors are again found largely at the element boundaries, as expected.

\subsubsection{Stress prediction in a $400\times$\ larger geometry}
\begin{figure}[h!]
    \centering
    \includegraphics[width=\columnwidth,keepaspectratio]{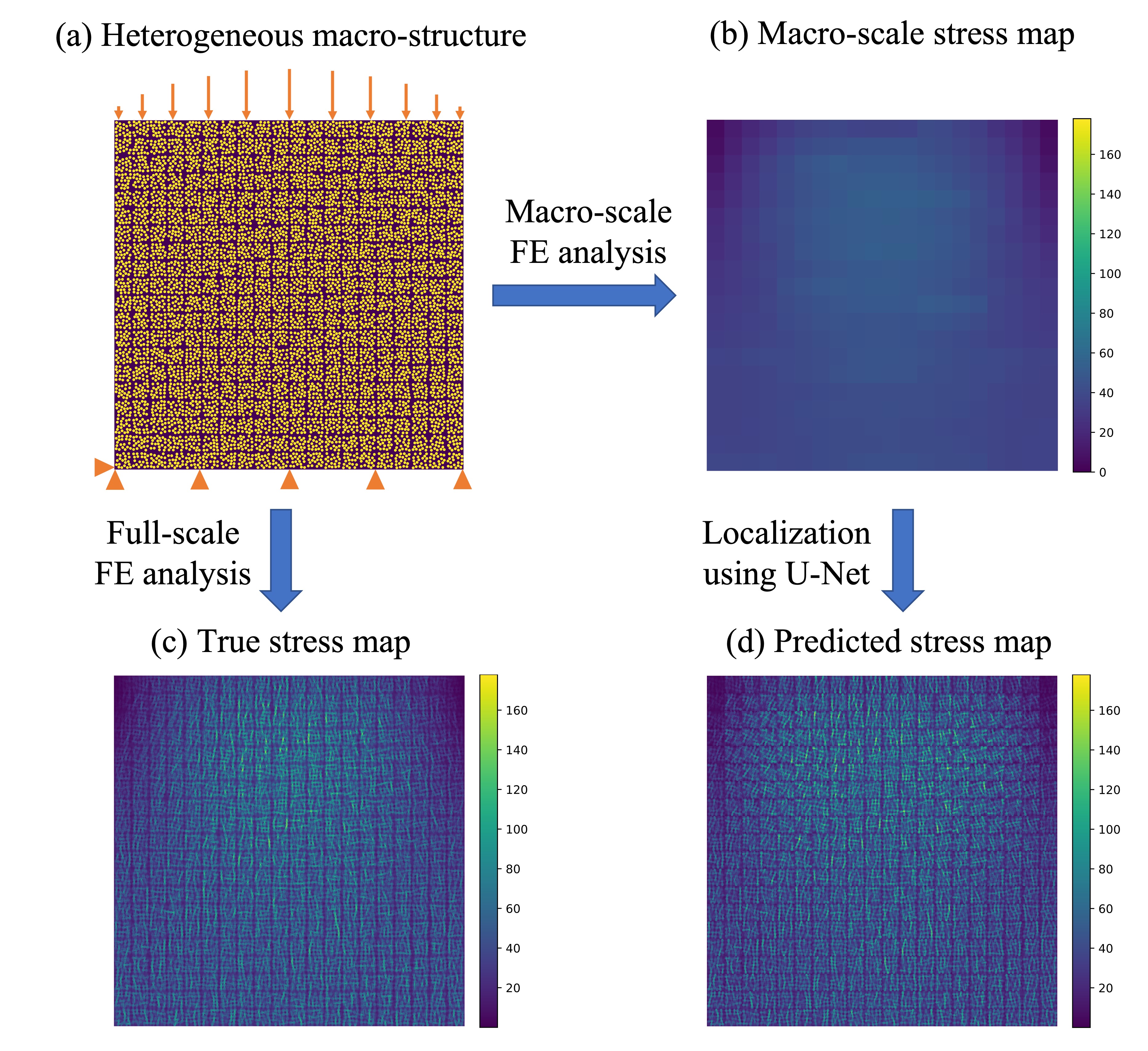}
    \caption{Multiscale analysis of a square domain of size $400\times$\ larger than the original microstructure subjected to $0.1\%$ tensile strain loading along the x-direction. (a) The boundary condition applied to the heterogeneous macro-structure. The von Mises stress maps resulting from: (b) macro-scale FE analysis of the $20\times20$ discretized square domain, (c) the subsequent ML-driven multiscale analysis, and (d) the full-scale FE analysis.}
    \label{multiscale_vonMises_20x20}
\end{figure}

A multiscale analysis of a square domain of size $400\times$\ larger than the original microstructure is performed as shown in Figure \ref{multiscale_vonMises_20x20}(a). The square domain is subjected to a parabolic stress loading on the top edge while keeping the bottom edge fixed as illustrated in Figure \ref{multiscale_vonMises_20x20}(a). The maximum stress is $64 MPa$ applied at the mid-point of the top-edge of the domain. This square domain is discretized by $20\times20$ square FE elements. A $20\times20$ homogenized macro-scale model is built by performing homogenization using the ML-model. Figure \ref{multiscale_vonMises_20x20}(b) shows the average von Mises stress in this $20\times20$ square domain from the macro-scale plane strain FE analysis.  Figure \ref{multiscale_vonMises_20x20}(c) and \ref{multiscale_vonMises_20x20}(d) show the predicted von Mises stress maps resulting from the full-scale FE analysis and the ML-driven multiscale analysis, respectively. These plots show that the predicted stress maps captures the local stress concentrations in the microstructure. Furthermore, the ML-driven multiscale analysis is remarkably fast as it takes on the order of seconds to analyze the large macro-structure, which is orders of magnitude faster when compared to the hours required by homogenization based $FE^2$ analysis and days required by the full-scale FE analysis on the same computational platform.

\section{Conclusions} \label{sec:conclusions}
This paper describes a novel approach to perform multiscale mechanics modeling in a deep learning framework. The proposed ML-driven multiscale analysis approach is based on an ML-model that predicts the local stress tensor fields in an elastic fiber-reinforced composite microstructure. The ML-model is trained to predict stress tensor fields resulting from a separate tensile and shear strain loading. This ML-model, even though trained only on a few images, shows excellent agreement with the reference solution obtained from finite element (FE) analysis. In addition, the ML-model accurately predicts micro-scale stresses under a multiaxial strain state, by applying the superposition principle for a linear elastic material. This pretrained ML-model is massively more efficient than FE analysis at both upscaling (homogenization) and downscaling (localization). Therefore, this model achieves remarkable increases in efficiency in the proposed ML-driven multiscale analysis approach. The ML-driven multiscale analysis approach shows good agreement with the corresponding full-scale FE analysis, with orders of magnitude less effort. Furthermore, the approach is shown to be applicable to a variety of macro-structure sizes, loadings, and boundary conditions. Finally, this work demonstrates that the challenge presented by limited quantities of data in mechanics modeling is in part compensated by the richness of information present in each materials datum. 

\indent The current approach expedites multiscale analysis of heterogeneous elastic 2-phase materials, but the framework is far more generalizable and inspires further extension to more complex materials and applications.  One possible advantage of this image based ML-model is that it can be trained directly on data from experimental imaging techniques such as X-ray-tomographic scans. While only a 2D approach is presented in this work, the very low computational cost of the deep-learning approach and its ability to learn from a very small amount of data also makes it an obvious candidate for 3D applications by using a 3D U-Net. Another future direction is to address more complicated mechanics, such as interfacial debonding, plasticity, and damage, which makes the problem history-dependent. The challenge in such models will be to accurately predict the evolution of micro-scale stress fields with time. 
Broader applications of the proposed approach include efficient multiscale analysis of complex materials for uncertainty quantification, multiscale design, and topology optimization. Beyond mechanical problems, this framework can be applied to several other multiscale physical phenomena, such as heat transfer, fluid flow, and coupled thermo-mechanical problems.

\section*{Acknowledgements}
Research was sponsored by the Army Research Laboratory and was accomplished under Cooperative Agreement Number W911NF-12-2-0023, \\ W911NF-12-2-0022 and W911NF-22-2-0014. The views and conclusions contained in this document are those of the authors and should not be interpreted as representing the official policies, either expressed or implied, of the Army Research Laboratory or the U.S. Government. The U.S. Government is authorized to reproduce and distribute reprints for Government purposes notwithstanding any copyright notation herein.

\bibliographystyle{elsarticle-num} 
\bibliography{cas-refs}

\end{document}